\documentstyle[aps,prb,epsf,multicol]{revtex}

\begin{document}
\setlength{\topmargin}{-1.5cm}
\newcommand\tJ{{$t-J$}}
\newcommand{\Secref}[1]{Section~\ref{sec:#1}}
\newcommand{\Eqnref}[1]{Eq.~(\ref{eq:#1})}
\newcommand{\Figref}[1]{Fig.~\ref{fig:#1}}
\newcommand{\utrans}[1]{U_{\widehat{\Omega}_{#1}}}
\newcommand\hc{{\hbox{\rm h.c.}}}
\newcommand\im{{\hbox{\rm Im}}}
\newcommand\re{{\hbox{\rm Re}}}
\newcommand\br{{\mathbf r}}
\newcommand\brp{{\mathbf r}'}
\newcommand{\spin}[1]{{\mathbf S}_{#1}}
\newcommand\half{{\frac{1}{2}}}
\newcommand\quarter{{\frac{1}{4}}}
\newcommand\etal{{\em et al.}}



\title{Staggered flux and stripes in doped  antiferromagnets}

\author{Martin Andersson and Stellan \"Ostlund}

\address{Department of Theoretical Physics and Mechanics, \\ 
	Chalmers University of Technology and G\"{o}teborg University, \\ 
	S-412 96 G\"{o}teborg, Sweden}

\date{\today}

\maketitle



\begin{abstract}
We have numerically investigated whether or not a mean-field theory
of spin textures  generate fictitious flux in the doped two dimensional 
$t-J$-model.  First we consider the properties of uniform systems and then
we extend the investigation to include models of striped phases where a 
fictitious flux is generated in the domain wall providing a possible source
for lowering the kinetic energy of the holes. We have compared the energetics
of uniform systems with stripes directed along the (10)- and (11)-directions of
the lattice, finding that phase-separation generically turns out to be
energetically favorable. In addition to the numerical calculations, we
present topological arguments relating flux and staggered flux to geometric 
properties of the spin texture. \cite{Ostlund01} The calculation is based on a
projection of the electron operators of the $ t-J $ model into a spin 
texture with spinless fermions.

\bigskip
\noindent
PACS number(s): 75.10.Jm
\end{abstract}


\begin{multicols}{2}
\narrowtext



\section{Introduction}
\label{sec:qhe}

It is well known that topological spin-textures are important in quantum Hall 
ferromagnets.~\cite{Girvin99,Girvin00} It is also known that an 
electromagnetic flux through a system of tight binding electrons on a lattice 
can lower the electronic kinetic energy.~\cite{Hasegawa89} Topological 
arguments also suggest that a possible source of this flux could be the 
formation of a spin-texture, which would suggest a relationship between doping
and magnetic flux.~\cite{Wen89} However, for the two dimensional 
antiferromagnetic Heisenberg model it has been 
argued~\cite{Haldane88,Einarsson91} that, taking the continuum limit and
looking at long wavelength fluctuations about the N{\'e}el state, there is no
topological term in the effective action. Although this argument is
correct, it does not, however, answer the question whether or not 
spin-textures can be important on a length scale comparable with the lattice.
Furthermore it does not address the issue of second-neighbor 
hopping.  These ideas will be explored in the present paper which extends ideas
presented previously.~\cite{Ostlund01} In particular, we generalize the 
topological arguments given in Ref.~[\onlinecite{Ostlund01}] and present a
numerical comparison of the energy of flux generating spin textures and 
flux free spin configurations for uniform systems. Besides uniform systems, we
also consider if spin- and charge-stripes arise naturally as a topological 
fictitious flux generating spin texture.

The paper is organized as follows: In \Secref{tJmodel} we introduce the 
{\tJ}-model and provide some background material on its properties. We proceed
in \Secref{adiabatic} by deriving the effective model which we will work with.
This model turns out to include topological fluxes which are discussed in
\Secref{fluxproperties} and in \Secref{fermionswithflux} we review the effect
of such a flux on a system of free electrons. \Secref{uniform} contains a
numerical mean-field investigation of the energetics of the system, comparing
flux generating states and more regular spiralling states where there is no
flux generated. In \Secref{stripes} we extend the discussion of uniform systems
to include stripes forming antiphase domain walls between N{\'e}el ordered 
regions. We describe of stripe model and present data from numerical 
calculations comparing different stripes. Finally, we summarize our results in
\Secref{conclusion}.



\section{The $t-J$ model}
\label{sec:tJmodel}
In order to explore these ideas we employ the {\tJ} model which most simply 
captures the competition between Heisenberg exchange and kinetic energy:
\begin{equation}
H = \sum_{\langle\br\brp\rangle}
	\left[ -t\left( c_{\br\sigma}^{\dagger}c_{\brp\sigma}
	+ \hbox{h.c.}\right) +J \left( \spin{\br}\cdot\spin{\brp}-
	\quarter n_{\br}n_{\brp}\right)\right].
\label{eq:tjham}
\end{equation} 
The summation is restricted to nearest-neighbor pairs on the square
lattice and the spin operator is given by
$\spin{\br}=\half c_{\br\alpha}^{\dagger}
\vec{\sigma}^{\alpha\beta}c_{\br\beta}$, where $\vec{\sigma}=(\sigma_x,
\sigma_y,\sigma_z)$ is the vector of Pauli matrices. All states containing
doubly occupied sites have been excluded from the Hilbert space, leaving
three states per site: $|\hspace{2mm}\rangle$, $|\uparrow\rangle$, and 
$|\downarrow\rangle$.  A natural generalization, deferred to later in this 
paper, is to add Coulomb-repulsion between particles occupying 
nearest-neighbor sites. 

Striped phases have been found experimentally in high-$T_c$ materials
to which the present model has been applied. There is an ongoing
debate regarding the existence of stripes in the {\tJ}- and Hubbard
models. Stripes were first found in Hartree-Fock solutions of the Hubbard
model~\cite{Zaanen89}, but the stripes found in these calculations had
one hole per unit length of the stripe, in contrast to the results from
experiments where half a hole per unit stripe length is found. From
DMRG calculations on finite systems (of the order
$20\times 10$). White and Scalapino~\cite{White98:2,White98:3,White00:2}
find stripes in a wide range of dopings. For instance, using $J=0.35t$
they find stripes for dopings in the interval $0<x<0.3$. For $x<0.125$
the stripes have half a hole per unit cell of stripe, in agreement
with experiments, and the distance between two consecutive stripes is
$d=(2x)^{-1}$. For higher dopings they find that there is one hole
per unit cell of the stripe and that the inter-distance between the
stripes is $d=x^{-1}$. On the other hand, using quantum Monte-Carlo
calculations Hellberg and Manousakis~\cite{Hellberg97,Hellberg99}
find that uniform or phase-separated states are energetically
favorable. In this case, the formation of stripes would also require
the existence of a long-range Coulomb interaction preventing an ordinary
phase-separation~\cite{Kivelson96,Low94}.

Incommensurate states were discussed in connection to the {\tJ} model
before the notion of stripes was introduced.  It was found by Shraiman and
Siggia~\cite{Shraiman88,Shraiman89} using a continuum limit of the model,
that the antiferromagnetic order of the undoped {\tJ} model is unstable
against the formation of a spiral state for small dopings. Using various
mean-field approaches other authors~\cite{Jayaprakash89,Schulz90,Kane90}
came to similar conclusions.  We will return to the spiral instability in
\Secref{uniform} using the effective model to be described in the next
section. It will be shown how a small twist in the spin-order leads to
a reduction of the kinetic energy of the order $tx$ while the loss in
exchange energy is of the order $Jx^2$; showing that for small enough
dopings there is energy to be gained by twisting the spin-order.

In addition to spin textures, Affleck and
Marston~\cite{Affleck88,Marston89}  discussed the possibility of
flux textures.  They replaced the two spin components of the electrons
by $N$ different flavors, extending the SU(2) spin symmetry to SU($N$).
Taking the limit $N\rightarrow\infty$, they obtained an essentially exact
mean-field model to which they numerically looked for solutions.
In particular they found a phase, called the flux phase, where the sum of the phases
of the link-operators $\chi_{\br\brp}= c_{\br}^{\dagger}c_{\brp}$
around a plaquette equals $\pm\pi$.  This is interpreted as half a flux
quantum penetrating each plaquette. These phases do not come from a real
electromagnetic field and are therefore referred to as fictitious.

Work by Hasegawa {\etal}~\cite{Hasegawa89} showed that the energy of
non-interacting spinless fermions has a minimum when a uniform flux,
corresponding to one flux quantum per particle, threads the system. States
having $\Phi =n$ (in units of the flux quantum) are referred to as
commensurate flux states. This shows that a fictitious flux can lower
the kinetic energy of the particles. The commensurate flux states have
also been considered in connection with the {\tJ} model.

Another possibility for a flux state is to have a staggered flux through
the system. In the case of half a flux quantum per plaquette there is no
difference between uniform- and staggered fluxes, so the Affleck-Marston
state can be thought to belong to this category as well. Inspired by
the work of Shraiman and Siggia, Kane {\etal}~\cite{Kane90} suggested
a double spiral state showing a staggered chiral spin order, and hence
also, according to a result of Wen {\etal}~\cite{Wen89}, a staggered fictitious
flux. In a staggered flux state, the time-reversal symmetry can be
broken locally but not globally, as the system is invariant under a
time-reversal operation followed by a lattice translation, just like
a N{\'e}el state. Staggered flux phases have also been investigated by
other groups~\cite{Harris89,Poilblanc90,Barford91,Hsu91}. The effective
model used by Barford and Lu~\cite{Barford91} coincides with the model
derived in the next section. 

Our paper expands on results found by previous authors; we find that
certain spin textures and charged stripes in particular are coupled by 
the creation of a fictitious $ \pi $ flux which we show is a 
natural consequence of a stripe with broken rotational symmetry.



\section{Deriving an effective model}
\label{sec:adiabatic}

In order to make progress, we make certain simplifications of the {\tJ} model.
First, following Schulz~\cite{Schulz90}, we introduce a local quantization 
axis $\widehat{\Omega}_{\br}$ at site $\br$. In terms of spherical coordinates
we write $\widehat{\Omega}_{\br}=(\sin\theta_{\br}\cos\phi_{\br},
\sin\theta_{\br}\sin\phi_{\br},\cos\theta_{\br})$. This local SU(2) 
transformation on $c_{\br}$, denoted by $\utrans{\br}$, must fulfill the 
equation
\begin{equation}
\utrans{\br}\sigma_z\utrans{\br}^{\dagger} = \widehat{\Omega}_{\br}
\cdot\vec{\sigma}.
\end{equation}
As can be seen from the above equation, specifying 
$\widehat{\Omega}_{\br}$ determines $\utrans{\br}$ only up to a rotation about
the new local $z$-axis. For example, we may choose our SU(2) transformation 
according to
\begin{equation}
	\utrans{\br} = 	\exp\left[ -i\frac{\theta_{\br}}{2}
			\widehat{\omega}_{\br}\cdot
			\vec{\sigma}\right] ,
\end{equation}
where $\widehat{\omega}_{\br}=\widehat{(\widehat{z}\times\widehat{\Omega}
_{\br})}=(-\sin\phi_{\br},\cos\phi_{\br},0)$.

Expressing the {\tJ} Hamiltonian in terms of this new spin-coordinate system
we find
\begin{eqnarray}
\label{eq:tjgen}
H = \sum_{\langle \br\brp\rangle }&\Bigl[ & -t\left( 
	c_{\br\alpha}^{\dagger}M_{\alpha\beta}^{\br\brp}
	c_{\brp\beta}+\hc \right) \cr
	&&+ J\left( 
	S_{\br}^{\alpha}S_{\brp}^{\beta}Q_{\alpha\beta}^{\br\brp}
	-\quarter n_{\br}n_{\brp}\right)\Bigr] ,
\end{eqnarray}
with
\begin{eqnarray}
M^{\br\brp} 	&=&	(\utrans{\br})^{\dagger}\utrans{\brp} \cr
Q^{\br\brp} 	&=&	R_{\widehat{\Omega}_{\br}}^{-1}
			R_{\widehat{\Omega}_{\brp}} \cr
(R_{\widehat{\Omega}})_{ij}
	      	&=&	\cos\theta\delta_{ij}+(1-\cos\theta )\omega_i\omega_j
			-\sin\theta\epsilon_{ijk}\omega^k .
\end{eqnarray}
We note that $R_{\widehat{\Omega}}$ is the SO(3) rotation-operator induced by 
the SU(2)-transformation $\utrans{}$.

Thinking of the {\tJ} model as the large-$U$ limit of the Hubbard model, we
know that there is a gap between the Hubbard bands scaling as $U$, 
corresponding to the energy cost for a double occupancy. Following 
Schulz~\cite{Schulz90}, 
who neglected holes in the lower Hubbard band, we will throw away the upper 
Hubbard band because of this large gap when we consider hole-doping, 
corresponding to the removal of states containing double occupancies from the 
Hilbert space of the Hubbard model.
Since the quantization axis at a site is locally determined by 
$\widehat{\Omega}_{\br}$, we
can arbitrarily assume that the upper Hubbard band is associated with spin
down relative $\widehat{\Omega}$. Hence, our effective model is obtained by
keeping only the terms in \Eqnref{tjgen} associated with spin-up particles.
The spin of an electron at site $\br$ 
will now be determined by the field $\widehat{\Omega}_{\br}$. As the simplest 
approximation, we will consider the $\widehat{\Omega}$-field as a classical 
field, neglecting spin-fluctuations in the system. Keeping only terms
containing particles aligned with the positive local $z$-axis we obtain an 
effective Hamiltonian
\begin{equation}
H_{\rm eff} = \sum_{\langle\br\brp\rangle}\left[ 
	-(\tau^{\br\brp}c_{\br}^{\dagger}c_{\brp}+\hc )
	+K^{\br\brp}n_{\br}n_{\brp} \right]
\label{eq:tjeff}
\end{equation}
with $\tau^{\br\brp}=t M_{11}^{\br\brp}$, 
$K^{\br\brp}=\quarter J (\widehat{\Omega}_{\br}\cdot
\widehat{\Omega}_{\brp}-1)$, and $c_{\br}=c_{{\br}\uparrow}$. This Hamiltonian
describes a system of spinless fermions moving in a lattice with position 
dependent hopping amplitudes and interaction strengths.\cite{Barfoot}

The effects of the spin texture on the charge-motion is therefore to
generate an effective hopping amplitude, $t\longrightarrow t\cos\frac{\theta}
{2}$ and the appearance of a fictitious magnetic field. We also note that 
there is a Coulomb like nearest-neighbor interaction of strength 
$\quarter J(\cos{\theta}-1)$. When $\cos\theta < 1$ this leads to an 
effective attraction between particles, which hints that the system may favor 
a phase-separation when being doped.

We have already mentioned that there is a degree of freedom in $U_{\widehat{\Omega}}$ not 
being fixed by $\widehat{\Omega}$. Since the effect of a spin-rotation 
about the local $z$-axis on the ``up'' spin only introduces a phase factor,
it will be indistinguishable from a local electromagnetic gauge transformation
in our approximation. Hence, the set of physically 
inequivalent choices of $\utrans{}$ are determined by $\widehat{\Omega}$, i.e. 
they belong to SU(2)/U(1)$\cong S^2$.

\begin{figure}[tbh]
\centerline{\epsfxsize=0.8\columnwidth\epsffile{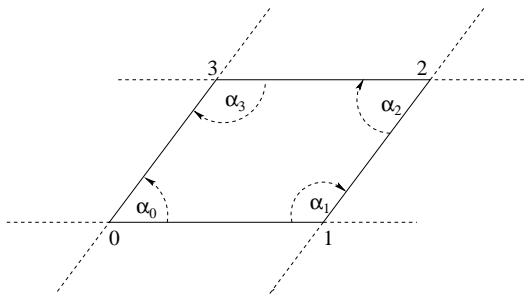}}
\vspace{0.2cm}
\caption[spinmap]
{\label{fig:spinmap}
        The mapping of a real-space plaquette into
        spin space. The solid angle spanned by the mapping is given by
        $\Sigma = \sum_i \alpha_i-2\pi$.
}
\end{figure}

The physical degrees of freedom of the hopping are contained within the size of
the hopping amplitude and the gauge invariant parts of the complex phases of 
the hopping elements. In case of nearest-neighbor hopping only, the smallest 
closed loop that can be formed is around a plaquette in the lattice, see
\Figref{spinmap}, and the
flux enclosed by such a counter-clockwise path 
$0\rightarrow 1\rightarrow 2\rightarrow 3 \rightarrow 0$ is given by
\begin{equation}
\label{eq:flux}
	\Phi_{0123} = {\rm Im}\ln (\tau^{01}\tau^{12}\tau^{23}\tau^{30}).
\end{equation}
One may show that this flux $\Phi_{0123}$ is equal to half the solid angle
enclosed by the shortest path on the spin sphere connecting the points
$\{\widehat{\Omega}_i\}_{i=0}^3$. Thus the flux is equal to $2\pi Q$, where
$Q$ is the topological charge represented by the plaquette.

In the following discussion we will refer to the flux as a fictitious flux, 
in contrast to a ``real'' electromagnetic flux that would come from an applied
magnetic field. The reason for this distinction is
that the fictitious spin generated flux is only seen by the spinless fermions
in the system and is unrelated to a physical electromagnetic field. 
Furthermore, the fictitious flux does not couple to the charge of the 
fermions, but rather to the $z$-component of the spin measured in the local 
spin-coordinate system. Since all particles in our system are polarized along
the positive $z$-axis, they will appear as having the same 
fictitious charge. However, the flux can still drive currents through the
system and in principle it is possible for the fictitious flux to 
cancel the effect of an external electromagnetic flux on the particles in the 
system. 

A physical magnetic field giving one flux quantum per plaquette is enormous. 
If we assume that the lattice constant is $a\simeq5$~{\AA}, the resulting 
magnetic field is $B=h/(ea^2)\simeq 10^4$~T.  This energy is much larger than 
the typical electronic energies per site, being of the order a few eV.



\section{Properties of fictitious fluxes}
\label{sec:fluxproperties}

In Ref.~[\onlinecite{Ostlund01}] we investigated the relations between the 
magnitude of 
the hopping $\tau^{\br\brp}$ and its complex phases for both ferromagnetic and 
antiferromagnetic spin-configurations. In the present section we shall briefly
review and generalize these arguments. 

We again consider a square lattice with a set of spins placed at each lattice
site $\br$. The interior angles on the surface of the sphere are
described by angles $\alpha_{i} $ as can be seen in \Figref{spinmap}.
The fictitious flux through the 
plaquette is equal to half the solid angle covered by the plaquette in 
spin space, which by spherical geometry is given by  the sum
of the interior angles in excess of $ 2 \pi$. It is obvious from 
\Figref{spinmap} that if $\theta_{{\br\brp}}$, the
angle subtended by the arcs on the sphere  is small, the
area of the spherical parallelogram, and hence the fictitious flux, will be
small as well. The following expressions give the size of the hopping and the 
fictitious flux through the plaquette: 
\begin{eqnarray}
\label{eq:fmflux}
	|\tau^{\br\brp}| &=& t\cos\frac{\theta}{2} \cr
	\Phi_{0123}	 &=& \half\sum_{i} \alpha_i  - \pi 
\end{eqnarray}
For small values of  opening angles on the spin sphere 
$\theta_{\br\brp}\simeq\theta $  and 
$ \alpha_i \simeq \alpha_{2+i}  $ the  flux is approximately given by
$\Phi_{0123}\simeq\frac{\theta^2}{2}\cos\frac{\alpha_1 -\alpha_2 }{2}$, showing
that the flux is bounded by $|\Phi|\leq\frac{\theta^2}{2}$.

\begin{figure}[tbh]
\centerline{\epsfxsize=0.6\columnwidth\epsffile{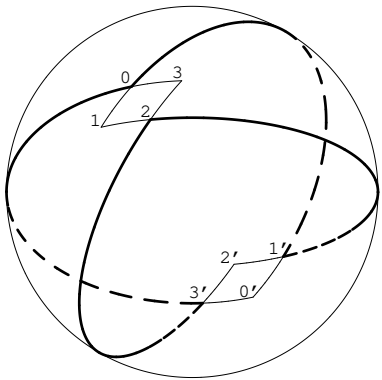}}
\vspace{2mm}
\caption[afmsphere]
{\label{fig:afmsphere}
        This figure shows the path taken on the spin sphere when going
        around a plaquette in an antiferromagnetic spin-configuration.
}
\end{figure}

If we instead turn our attention to antiferromagnetic configurations described
by $\theta\geq\frac{\pi}{2}$, the situation changes drastically. The path taken
in spin space when going around a plaquette in an antiferromagnetic 
configuration is shown in \Figref{afmsphere}. If we denote the 
antiferromagnetic (staggered) spin at site $\br$ by $\widehat{\Omega}_{\br}$,
the path taken is $\widehat{\Omega}_{0}\longrightarrow
\widehat{\Omega}_{1}'\longrightarrow\widehat{\Omega}_{2}\longrightarrow
\widehat{\Omega}_{3}'\longrightarrow\widehat{\Omega}_{0}$, where 
$\widehat{\Omega}'=-\widehat{\Omega}$ denotes the antipodal point on the
sphere. 
Redefining $\theta$ to denote the opening angle between two neighboring
staggered spins, we find the following relationships:
\begin{eqnarray}
\label{eq:afmflux}
	|\tau^{\br\brp}| &=& t\sin\frac{\theta}{2} \cr
	\Phi_{01'23'}	 &=& \pi - \half \sum_i (-1)^{i} \alpha_i
\end{eqnarray}
We note that the size of the fictitious flux is now completely
decoupled from the opening angle $\theta$, as long as 
$\theta\leq\frac{\pi}{2}$. Furthermore, in the antiferromagnetic case the
fictitious flux is staggered since the path on a neighboring plaquette will
be traversed with the sublattices exchanged. 

If we allow for the possibility of next-nearest-neighbor (diagonal) hopping,
we have to take into account additional gauge invariant fluxes. There are four
of these for each square plaquette, defined by the removal of one of the four
vertices of the plaquette. In Ref.~[\onlinecite{Ostlund01}] we show that there
is a topological constraint relating these four fluxes. This constraint takes the 
form
\begin{equation}
\Phi_{01'2}+\Phi_{023'}-\Phi_{01'3'}-\Phi_{1'23'}=2\pi n,
\label{eq:topologicalconstraint}
\end{equation}
where $n=0$ for a ferromagnet and $n=\pm 1$ for an antiferromagnet. In case of
a ferromagnet a prime does not denote the antipodal point, but rather the
point itself. This 
relation is easy to verify by looking at \Figref{afmsphere} noting that the
sphere is exactly covered by the four regions in 
\Eqnref{topologicalconstraint}. The topological
constraint, \Eqnref{topologicalconstraint}, does not rely on the assumption
of a spherical parallelogram and does also hold in the presence of an external
electromagnetic flux. Counting degrees of freedom, we know that there are 
two degrees of freedom per plaquette (or site) in choosing the 
spin-configuration and in addition we have one parameter coming from an
external flux. All in all, there are three free parameters per plaquette and
hence we expect that the four fluxes through the sub-triangles are related by
a single constraint, given above.



\section{Spinless free fermions with flux}
\label{sec:fermionswithflux}

Before turning to a more through investigation of the physics of the
effective Hamiltonian, \Eqnref{tjeff}, we will discuss the effects of a flux
through a system of free spinless fermions. Hasegawa \etal~\cite{Hasegawa89}
investigated a system of free electrons on a square lattice with a uniform 
magnetic flux. They found that the energy is minimized when there is 
exactly one flux quantum per particle, i.e. the optimal flux per plaquette 
is related to the
doping according to $\Phi =(1-x)\Phi_0$, where $\Phi_0$ is the flux quantum. 
To illustrate this effect we have diagonalized the Hamiltonian for such
a system of free fermions in a uniform flux $\Phi$ for different values of
$\Phi$, finding the single-particle energies $\epsilon_i(\Phi)$. The total
energy of the system is then found by summing up the single-particle energies
according to 
\begin{equation}
E(\Phi ,n)=\sum_{\epsilon_i(\Phi)<\epsilon_F(n,\Phi)}\epsilon_i(\Phi),
\end{equation}
where $\epsilon_F(n,\Phi)$ is the Fermi energy corresponding to filling 
$n$ and flux $\Phi$.
\begin{figure}[tbh]
\centerline{\epsfxsize=0.9\columnwidth\epsffile{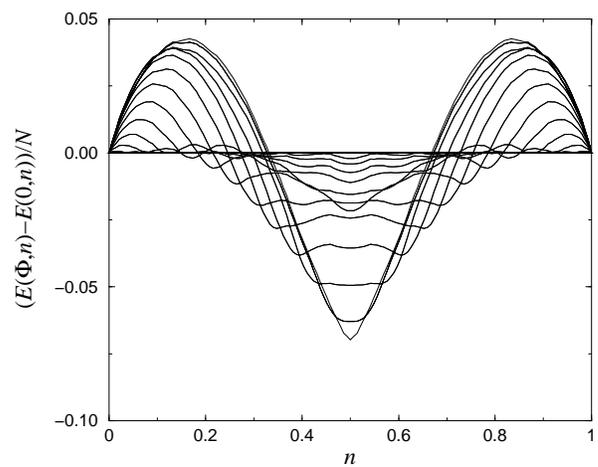}}
\caption[freeuniform]
{\label{fig:freeuniform}
        This figure shows energy versus filling for free electrons using a set
        of different uniform fluxes, $\Phi =q\pi /12$,
        where $q=0,1,\dots ,12$. In the center of the figure, i.e. at $n=1/2$,
        the optimal energy is given by $\Phi=\pi$. Moving to the right, the
        next minima corresponds to $\Phi=11\pi/12$ and so on, finally finding
        an optimal flux $\Phi=0$ at $n=1$.
}
\end{figure}
In \Figref{freeuniform}, we plot $E(\Phi ,n)-E(0,n)$, i.e. the energy per site
for different fluxes compared to the flux free case. The figure
clearly shows how the optimal flux changes with doping. We also note that the
system is particle-hole symmetric and hence $E(\Phi ,n)=E(\Phi,1-n)$.

\begin{figure}[tbh]
\centerline{\epsfxsize=0.9\columnwidth\epsffile{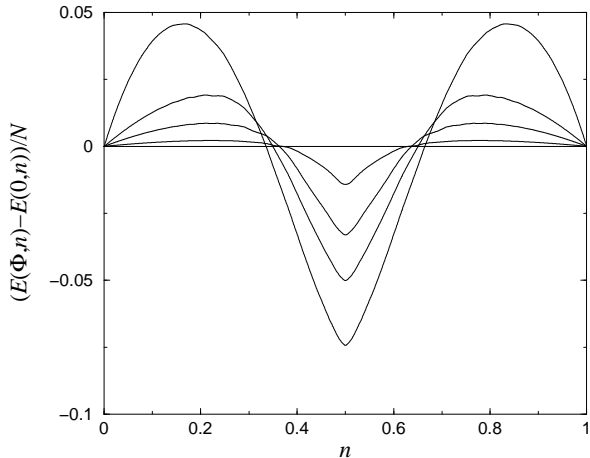}}
\caption[freestaggered]
{\label{fig:freestaggered}
        This figure shows energy versus filling for free electrons compared to
        the flux-free case for a set
        of different staggered fluxes, $\Phi =q\pi$. Going upwards in the
        center of the figure $q=1,1/2,1/3,1/6$. Note that for a certain
        doping, the minimum energy is obtained either for $\Phi =0$ or
        $\Phi =\pm\pi$. The crossing points occur at $n\simeq 1/2\pm
        0.165$.
}
\end{figure}

However, the flux that is generated by the antiferromagnetic skyrmions is
staggered in which case the Hamiltonian can be exactly diagonalized. 
The spectrum of this system, assuming $\theta_x=\theta_y=\theta$, is
\begin{eqnarray}
\epsilon_{\mathbf k}(\Phi) &&= \pm 2t\sin\left(\frac{\theta}{2}\right)\cr
&&\times	\sqrt{\cos^2k_x+2\cos\frac{\Phi}{2}\cos k_x\cos k_y+\cos^2k_y}.
\end{eqnarray}
In \Figref{freestaggered} we show a plot of
the energy per site for different staggered fluxes. The figure shows that the
optimal flux is either $0$ or $\pm\pi$ depending on doping,  the only
choices which are consistent with time-reversal invariance.

From this analysis we conclude that it is reasonable when searching for
minimum energy spin textures to consider configurations supporting flux
$0$ (coplanar configurations) and $\pm\pi$.  In the next section we
will construct a mean-field theory based on these observations.


\subsection{Second-neighbor hopping}

As we have seen in \Secref{adiabatic}, an antiferromagnetic spin-configuration
strongly suppresses the effective nearest-neighbor hopping on the square lattice.
However, a next-nearest-neighbor hopping is compatible with 
antiferromagnetic order. For the purpose of illustration let us
consider the following purely kinetic Hamiltonian describing spinless 
fermions 
\begin{equation}
\label{eq:tandtpham}
H_{t-t'}=-t\sum_{\langle\br\brp\rangle}\left[c_{\br}^{\dagger}c_{\brp}+
	\hc\right]-t'\sum_{\langle\langle\br\br ''\rangle\rangle}
	\left[c_{\br}^{\dagger}c_{\br ''}+\hc\right] ,
\end{equation}
where $\langle\langle\br\br ''\rangle\rangle$ denotes next-nearest-neighbor
pairs. We remark that the sign of the nearest-neighbor
hopping $t$ is irrelevant as it can be changed by the transformation
$c_{\br}\longmapsto (-1)^{\br}c_{\br}$. This transformation leaves the sign
of $t'$ unchanged, and this sign is important. Without loss of generality,
we assume $t=1$.

Another symmetry operation of interest is the particle-hole transformation
$c_{\br}\longmapsto (-1)^{\br}c_{\br}^{\dagger}$. Under this operation, the
sign of the nearest-neighbor hopping is unchanged while the sign of $t'$ is 
changed, showing that next-nearest-neighbor hopping breaks the particle-hole
symmetry. Furthermore, the number operator $n_{\br}\longmapsto 1-n_{\br}$ as
particles are mapped into holes. This symmetry was seen in 
Figures~\ref{fig:freeuniform} and \ref{fig:freestaggered}.

Let us define $E(n,t')$ as the energy per site in the ground state
of \Eqnref{tandtpham} with next-nearest-neighbor hopping $t'$. It is easy to
show that as long as $n<1/2$, $E(n,|t'|)<E(n,-|t'|)$ showing that for small
fillings the energy is lower for the positive $t'$-case. If we instead
consider the region $n>1/2$ the particle-hole transformation discussed above
immediately tells us that $E(n,-|t'|)<E(n,|t'|)$, showing that the negative
$t'$-case is favorable. At precisely half-filling, the energy is independent of
the sign of $t'$.

Assume that we consider the case $n<1/2$ and $t'<0$. Then, according to the
discussion above, we would gain energy if we could somehow change the sign
of $t'$. One way of accomplishing this would be to add a uniform flux through
the system, corresponding to one flux quanta per square plaquette. This flux
would not affect the nearest-neighbor hopping, but it would change the sign
of $t'$ and therefore lower the energy of the system. Barford and 
Kim~\cite{Barford91:2} generalized the result of Hasegawa {\etal} to include 
the $t-t'$ model above and found that in the thermodynamic limit, 
the kinetic energy is minimized by a flux corresponding to one flux quantum 
per site plus or minus one flux quantum per particle.



\section{Hartree-Fock theory of uniform phases of $H_{\rm eff}$}
\label{sec:uniform}

It has been recognized for some time~\cite{Shraiman88,Shraiman89} that
a plausible response of a Heisenberg antiferromagnet to doping, is to form
a spiral spin-wave where the N{\'e}el 
order-parameter rotates uniformly around a fixed spin-axis as one moves along 
a symmetry axis in the lattice. This together with the fact that the
doped electron gas favors a staggered flux close to half-filling, indicates
that a state containing an antiferromagnetic spin-texture that generates both 
a staggered fictitious flux and a spiral-like order could lead to an 
energetically favorable state. 

In the remaining sections of this paper, we will address two
related questions. First, we investigate the effective Hamiltonian, \Eqnref{tjeff}, 
looking for the spin-textures which provide the energetically most favorable uniform states. 
In particular, we are interested in whether or not the system chooses to incorporate fictitious 
fluxes. Furthermore, it is known that the {\tJ} model has a tendency to 
phase-separate. Concerning the thermodynamic stability of the spiral states,
Hu {\etal}~\cite{Hu94} found that in a Hubbard model, for small dopings, 
the spiral phase is unstable against phase-separation (for $U/t\gtrsim 10$) or 
domain wall formation (for $U/t\lesssim 10$). For larger dopings, there are
regions in the phase-diagram, located around  $U/t\simeq 10$, where the spiral
phases are thermodynamically stable. This indicates that, in the {\tJ} model, 
for small dopings the spiral state is not thermodynamically stable.
An interesting question is; if the {\tJ} model prefers a 
flux-phase in some region of parameter space, can this thermodynamically 
stabilize the system, preventing it from phase-separation? 

Inspired by the recent interest in striped phases, in \Secref{stripes} we
use our approach to model different domain walls between 
N{\'e}el-ordered regions. These domain walls have an appealing
structure as they provide a smooth implementation of the antiphase boundary
and at the same time provides the electrons in the doped channel with a
fictitious flux.

\subsection{MFT-formulation}

We now look for different uniform phases of the effective 
Hamiltonian given in \Eqnref{tjeff}. The coefficients $\tau^{\br\brp}$ and 
$K^{\br\brp}$ are now given by
\begin{eqnarray}
\tau^{\br\brp} &=& t\sin\frac{\theta_{\br\brp}}{2}
	\exp [i {\cal A}(\widehat{\Omega}_{\br},
              \widehat{\Omega}_{\brp},\widehat{z})/2]\cr
K^{\br\brp}    &=&   -\frac{J}{4}(1+\cos\theta_{\br\brp}),
\end{eqnarray}
where $\widehat{\Omega}_{\br}$ denotes the local 
staggered spin-orientation at site $\br$, and 
$\cos\theta_{\br\brp}=\widehat{\Omega}_{\br}\cdot\widehat{\Omega}_{\brp}$. 
With this definition $\cos\theta_{\br\brp}=1$ for a N{\'e}el state and 
$\cos\theta_{\br\brp}=-1$ for a ferromagnet, where $\br$ and $\brp$ are 
nearest neighbors.
${\cal A}(\widehat{\Omega}_{\br},\widehat{\Omega}_{\brp},\widehat{z})$ 
is the solid angle of the spherical triangle spanned by the vectors
$\widehat{\Omega}_{\br}$, $\widehat{\Omega}_{\brp}$, and $\widehat{z}$.

The approach we will use is a simple mean-field theory assuming a fixed 
spin-texture $\{\widehat{\Omega}_{\br}\}$, defined as the direction of the 
quantization axis, $\widehat{\Omega}_{\br}$. We will assume 
that $\theta_{\br\brp}=\theta_x$ when $\br$ and $\brp$ are nearest horizontal 
neighbors, and $\theta_{\br\brp}=\theta_y$ when they are nearest vertical 
neighbors. First of all we perform a standard mean-field decomposition of the 
Hamiltonian, allowing only for mean-fields carrying no charge and momenta zero
or ${\mathbf Q}=(\pi ,\pi )$. This results in the following Hamiltonian:
\end{multicols}
\widetext
\begin{eqnarray}
H_{\rm MF}&=&\sum_{{\mathbf k}\in {\rm BZ'}}\Psi_{\mathbf k}^{\dagger}
	\biggl\{ -2t\left[
        \left(\sin\frac{\theta_x}{2}\cos k_x+
        \cos\frac{\Phi}{2}\sin\frac{\theta_y}{2}\cos k_y\right)\sigma_3
	+\sin\frac{\Phi}{2}\sin\frac{\theta_y}{2}\cos k_y\sigma_2
        \right]\cr
&&\hspace{4mm}-\frac{J}{2N}\Bigl[
        (2+\cos\theta_x+\cos\theta_y)\Delta^0\openone
	-(1+\cos\theta_x)\cos k_x\Delta^3_{cx}\sigma_3
        -(1+\cos\theta_y)\cos k_y(\Delta^2_{cy}\sigma_2+\Delta^3_{cy}\sigma_3)
        \Bigr]\biggr\}\Psi_{\mathbf k}\cr
&&+\frac{J}{4N}\Bigl[(2+\cos\theta_x+\cos\theta_y)(\Delta^0)^2-(1+\cos\theta_x)
        (\Delta^3_{cx})^2
-(1+\cos\theta_y)\Bigl( (\Delta^2_{cy})^2+(\Delta^3_{cy})^2 
        \Bigr)\Bigr] ,
\label{eq:mftham}
\end{eqnarray}
\begin{multicols}{2}
\narrowtext
\noindent
where we have introduced a two-component vector
$\Psi_{\mathbf k} =(c_{\mathbf k},c_{{\mathbf k}+{\mathbf Q}})^t$, mixing 
momenta $0$ and ${\mathbf Q}$. We have introduced the Pauli matrices as a 
basis for the $2\times 2$ matrices coupling the $\Psi_{\bf k}$'s, although
we want to emphasize that they have nothing to do with spin in this
context. The sum over momenta is reduced to half the Brillouin zone, 
defined by ${\rm BZ'}=\{ |k_x|+|k_y|\leq \pi :-\pi\leq k_x,k_y< \pi \}$. 
Furthermore, we have only kept those four 
order-parameters \cite{footop} that turn out to be non-zero numerically. These four fields are defined through
\begin{equation}
	\left\{
	\begin{array}{rcl}
        \Delta^0 & = & 
        \sum_{{\mathbf k}\in {\rm BZ'}} 
                \langle\Psi_{\mathbf k}^{\dagger}\openone
                \Psi_{\mathbf k}\rangle \\
        \Delta^2_{cy} &=&
        \sum_{{\mathbf k}\in {\rm BZ'}} 
                \cos k_y\langle\Psi_{\mathbf k}^{\dagger}\sigma_2
                \Psi_{\mathbf k}\rangle \\
        \Delta^3_{cx} &=&
        \sum_{{\mathbf k}\in {\rm BZ'}} 
                \cos k_x\langle\Psi_{\mathbf k}^{\dagger}\sigma_3
                \Psi_{\mathbf k}\rangle \\\
        \Delta^3_{cy} &=&
        \sum_{{\mathbf k}\in {\rm BZ'}} 
                \cos k_y\langle\Psi_{\mathbf k}^{\dagger}\sigma_3
                \Psi_{\mathbf k}\rangle
	\end{array}
	\right. ,
\label{eq:mft}
\end{equation}
where the average $\langle\hbox{ }\cdot\hbox{ }\rangle$ denotes a thermal 
expectation 
value with respect to the Fermi-distribution of quasiparticles of 
$H_{\rm MF}$. The order-parameter $\Delta^0$ is 
simply the number of particles in the system, while the other three 
correspond to hopping induced through the term $n_{\br}n_{\brp}$ in the 
effective Hamiltonian, \Eqnref{tjeff}. In particular, we note that 
$\Delta^0$ and $\Delta^3_{cx/y}$ are diagonal and hence do not mix momenta
${\bf k}$ and ${\bf k+Q}$. On the contrary, $\Delta^2_{cx}$ does mix the two,
and therefore carries a momentum ${\mathbf Q}$. As can be seen from 
\Eqnref{mftham}, this term only exists in the kinetic term when there is a 
non-zero staggered flux which reduces the translational symmetry of the model.


\subsection{Instability towards spiral-order at low dopings}

Before turning to the numerical results, let us now discuss the 
electron gas in \Eqnref{tjeff} at low dopings, confining our 
discussion to coplanar spin configurations and neglecting exchange effects in
the Heisenberg term so that all order-parameters, except $\Delta^0$, in 
\Eqnref{mft} are zero.
We choose a spin-structure
$\widehat{\Omega}_{\br} \cdot \widehat{\Omega}_{\br+\widehat{x}} 
= \cos{ \theta_x } $ and  
$\widehat{\Omega}_{\br} \cdot \widehat{\Omega}_{\br+\widehat{y}} 
= \cos{ \theta_y }$, allowing for an asymmetry between the $x$- and 
$y$-directions. This yields a trivial
system which is exactly diagonalized. The total energy per site as a function 
of density $ n $ and $ \theta_{i} $ ($i=x,y$) is given by
\begin{eqnarray}
\label{eq:unihartree}
E_{\theta_x,\theta_y } ( n )  &=& 
	\frac{1}{N}\sum_{ \epsilon_{\mathbf k} \le \epsilon_F(n) } 
	\epsilon_{\mathbf k} \left( \theta_x , \theta_y \right) \cr &&-
        \quarter J n^2 \left( \cos{ \theta_x } + \cos{ \theta_y } + 2 \right)
\end{eqnarray}
where $ \epsilon_F(n) $ is the Fermi energy corresponding to density $ n $ 
and 
\begin{equation}
 \epsilon_{\mathbf k}(\theta_x,\theta_y) 
	= -2t \left[ 	\sin\frac{\theta_x}{2}\cos k_x  + 
			\sin\frac{\theta_y}{2}\cos k_y \right]  .
\end{equation}

\begin{figure}[tbh]
\centerline{\epsfxsize=0.8\columnwidth\epsffile{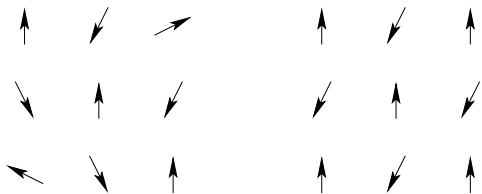}}
\caption[spiralvscanting]
{\label{fig:spiralvscanting}
        The difference between a (11) spiral state (left) and a canting state
        (right) is illustrated.
}
\end{figure}

Our description of the spin-order in terms of the $\theta$-angles does not
distinguish between spiral states and so called canted states. They both lack
fictitious flux and both have the same relative angle between 
nearest-neighbor spins. Only second neighbor terms resolve this degeneracy.
The difference 
between these two classes of states is illustrated in \Figref{spiralvscanting}.

If we make a series expansion of the energy per site in terms of the doping
$x$ we find the following expression to first order in $x$
\begin{eqnarray}
\label{eq:ensmalldop}
E_{\theta_x,\theta_y}(x) &=& 
                        - 2t\left(\sin\frac{\theta_x}{2}
			+\sin\frac{\theta_y}{2}\right) x -\quarter J(1-2x )\cr
 			&&\times(\cos\theta_x+\cos\theta_y+2)
                        + O(x^2).
\end{eqnarray}
Remember that we have made the transformation $ \theta_i \longmapsto 
\pi - \theta_i $, expressing the order relative to the antiferromagnet instead
of the ferromagnet. The energy is minimized by $ \theta_x = \theta_y =
2\arcsin [ 2tx / J ( 1 - 2 x ) ] \simeq 4tx /J$ 
for moderate dopings. The dependence on $tx /J$ clearly shows
the competition between the kinetic energy, which drives the system towards 
ferromagnetism, and the Heisenberg term, which favors antiferromagnetism. 
These results are consistent with those of Schulz~\cite{Schulz90}, who also
found this instability. Shraiman and Siggia~\cite{Shraiman88,Shraiman89},
using a more elaborate method, also found this instability, but their spiral
state has its pitch-vector along the $(10)$-direction rather than the 
$(11)$-direction as is
found here. In the Hubbard model, it is known from Hartree-Fock theory 
that the antiferromagnetic state is stabilized by the opening of a gap at the 
Fermi surface, and Schulz~\cite{Schulz90:2} has argued that a modulation of 
the spin-order along the $(10)$-direction opens up a gap more 
effectively than a spiral along the $(11)$-direction, i.e. it opens up a gap 
at a larger part of the Fermi surface. The first order
theory described in this section
does not take this fact into account. We
note that the deviation from N\'{e}el-order is proportional to the doping and
in the limit $J\rightarrow 0$, the deviation from N\'{e}el-order becomes
large for any finite doping. This is consistent with
Nagaoka's theorem~\cite{Nagaoka66}, stating that in the limit $J\rightarrow 0$,
a single hole doped into a Heisenberg antiferromagnet drives the system into a
ferromagnetic state.
The energy of the spiral state is given by 
$E=-\frac{4t^2x^2}{J(1-2x)}-J(1-2x)$ and hence we have for the 
second derivative of $E$ with respect to the filling $n$,
\begin{equation}
	\frac{\partial^2E}{\partial n^2}=-\frac{8t^2}{J(1-2x)^3}.
\end{equation}
We note that the energy versus filling is concave for dopings $x<1/2$,
showing that the spiral state has an instability towards phase-separation for all 
$J>0$. Recall that our expansion is only valid for small dopings, and in this
regime we expect the above conclusion to hold. Also, as mentioned earlier, we
cannot distinguish between the spiral- and canting states within this approach.
According to Kane {\etal}~\cite{Kane90} quantum fluctuations seem to stabilize
the spiral state compared to the canting state.

A similar analysis can be performed for fillings close to zero, i.e. $n\ll 1$.
In this case, measuring $\theta_i$ with respect to the ferromagnetic
configuration, the analogue of \Eqnref{ensmalldop} becomes (now keeping terms
up to second order in $n$)
\begin{eqnarray}
E_{\theta_x,\theta_y}(n) &=& 
                        - 2t\left(\cos\frac{\theta_x}{2}
			+\cos\frac{\theta_y}{2}\right) n\cr
			&&+2\pi tn^2
			\sqrt{\cos\frac{\theta_x}{2}\cos
			\frac{\theta_y}{2}} \cr &&
                  	+ \quarter Jn^2(\cos\theta_x+\cos\theta_y-2)
                        + O(n^3).
\end{eqnarray}
In particular we note how the second term introduces a coupling between the
spin-order in the $x$- and $y$-directions, allowing for an asymmetry between
$\theta_x$ and $\theta_y$. Minimizing the above energy with respect to the
angles $\theta_i$, we find that the ferromagnetic state is stable up to a
finite doping. At this point different things can happen depending on
$J/t$, the system
can pick a state where $\theta_x=0$ and $\theta_y=\pi$ (or vice versa), 
i.e. the system organizes itself ferromagnetically along the $x$-direction
while being an antiferromagnet along the $y$-direction. Another possibility 
is that the system chooses a pitch-vector along $(11)$ with 
$\theta_x=\theta_y=2\arccos\frac{(2-n\pi )t}{Jn}$.

The picture we have obtained is therefore that starting at zero filling, the
system remains in a ferromagnetically ordered state up to some threshold 
value of the 
filling. This threshold increases with decreasing values of the ratio $J/t$.
Above this threshold filling, the system can be a spiral spin-wave with pitch
along $(10)$ or $(11)$. When the filling approaches $n=1$, a
$(11)$ spiral state is optimal which continuously merges with the 
N\'{e}el-state as $n\rightarrow 1$. But we find in all cases that the system 
is unstable against phase-separation for small dopings.


\subsection{Numerics}

Given the thermodynamic instability of the spiral spin waves, we use the full
Hartree-Fock theory and take into account uniform phases that have a splay in
the N{\'e}el order-parameter. The aim
is to search the space of spin-textures, parametrized by $(\theta_x,\theta_y,
\Phi)$, to determine the one which minimizes the free energy of the system.
We will consider the following two types of spin-textures:
\begin{eqnarray}
&&\hbox{1. Coplanar states, described by $\Phi=0$.} \cr
&&\hbox{2. $\pi$-flux state, described by $\theta_x=\theta_y$, $\Phi=\pi$.}
\label{eq:spinconfig}
\end{eqnarray} 
The argument for only considering $\Phi=0$ and $\Phi=\pi$ states was given in
\Secref{fermionswithflux}; \Figref{freestaggered} showed that the energy
per site for a system of free fermions in a staggered flux is minimized by
either of these two choices.

Our numerical algorithms work within the grand canonical ensemble, 
with a free energy $G(T,\mu )=\langle H_{\rm MF}\rangle -TS-\mu N$, assuming a
fixed chemical potential $\mu$. Diagonalizing the mean-field Hamiltonian, 
\Eqnref{mftham}, we obtain a set of quasiparticle states specified by 
their momenta ${\mathbf k}\in {\rm BZ'}$, where ${\rm BZ}'$ is the reduced
Brillouin zone corresponding to $|k_x+k_y|\leq\pi$, and 
band-index. The band index refers to the two bands occurring because
of the staggered flux.
$\alpha\in\{ 1,2\}$. 
If we then minimize the free energy $G(T,\mu )$ with respect to the occupation
numbers $f_{{\mathbf k}\alpha}$ of the quasiparticles, we find that they are 
distributed according to the Fermi-distribution. The entropy, $S$, introduced 
above is defined through,
\begin{equation}
S=-k_B\sum_{\alpha,{\mathbf k}\in {\rm BZ'}}
	\Bigl( f_{{\mathbf k}\alpha}\ln f_{{\mathbf k}\alpha}
	+(1-f_{{\mathbf k}\alpha})\ln (1-f_{{\mathbf k}\alpha})\Bigr) ,
\end{equation}
where $\alpha\in\{ 1,2\}$ labels the two bands.

In the analysis of the numerical data we would rather consider the free energy
as a function of the number of particles $N$ than the chemical potential $\mu$.
This can be achieved by forming the Helmholtz free energy through the following
Legendre transformation; $F(N,T)=G+\mu N$. Having the Helmholtz energy,
we can use the Maxwell construction to discuss the thermodynamic stability of 
the phases.

Analyzing the Hartree-Fock theory involves the following procedure: 
Given a set of coupling constants, a spin-configuration, temperature $T$, and 
chemical potential $\mu$; pick a set of initial values of the mean-fields. 
Then solve for the quasiparticles of \Eqnref{mftham} and calculate the
new mean-fields using \Eqnref{mft}. The procedure is then iterated until the 
mean-fields have converged to a fixed point corresponding to a minimum in the
free energy. Since we are
interested in the spin-configuration minimizing the free energy $F(T,N)$, 
we will manually vary the spin-configuration parameters searching for a global
minimum of the free energy.

In practice, rather 
than choosing a certain value of the chemical potential, we choose a fixed 
filling, successively adjusting the chemical potential during the iterations. 
Since the chemical potential was assumed to be fixed during the derivation of 
the self-consistency equations, we have to make sure that the algorithm 
converges to the correct fixed point. We have checked explicitly in several 
cases that the correct fixed point is found. An advantage with this method
is that we can access all possible fillings, whether or not it is a 
thermodynamically stable region. This is not the case if we specify
the chemical potential, since the function $\mu(n)$ is not invertible.

\begin{figure}[tbh]
\centerline{\epsfxsize=0.9\columnwidth\epsffile{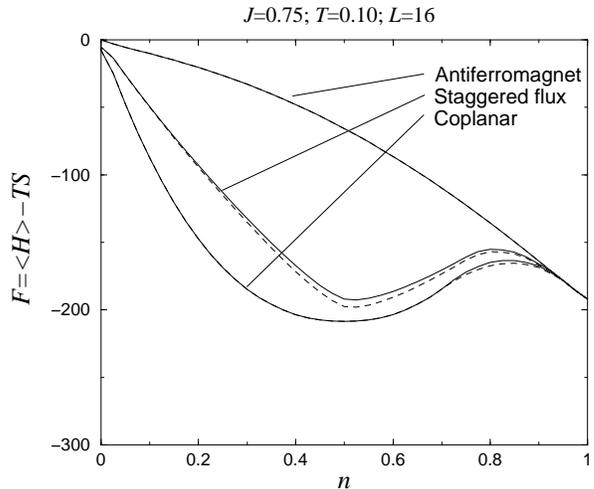}}
\caption[free1]
{\label{fig:free1}
        The Helmholtz free energy for a {\tJ} model with $J=0.75$ at
        temperature $T=0.1$ is shown for the coplanar, staggered-flux, and
        antiferromagnetic spin-configurations. Solid lines correspond to
        Hartree-Fock calculations, while the dashed lines correspond to
        Hartree calculations.
}
\end{figure}

\begin{figure}[tbh]
\centerline{\epsfxsize=0.9\columnwidth\epsffile{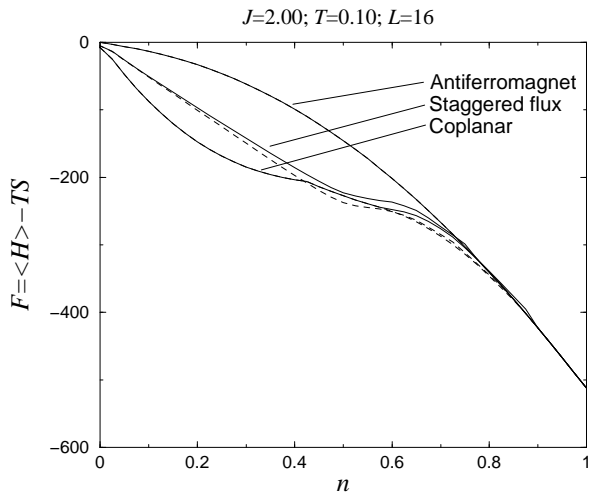}}
\caption[free2]
{\label{fig:free2}
        The Helmholtz free energy for a {\tJ} model with $J=2.00$ at
        temperature $T=0.1$ is shown for the coplanar, staggered-flux, and
        antiferromagnetic spin-configurations. Solid lines correspond to
        Hartree-Fock calculations, while the dashed lines correspond to
        Hartree calculations.
}
\end{figure}

Let us start by considering a simple Hartree-approximation, 
see Figures~\ref{fig:free1} and \ref{fig:free2}. This corresponds to putting 
$\Delta^2_{cy}=\Delta^3_{cx}=\Delta^3_{cy}=0$. Starting at zero temperature, 
we find that within the Hartree approximation there is a critical value 
$J_c\simeq 1$ of the coupling $J$ below which the coplanar phase always 
dominates over the flux-generating configurations.
When $J>J_c$, there appears a region around 
$n\simeq 0.5$ where the flux-states are the energetically lowest states, 
see \Figref{free2}. When the flux-state minimizes the free energy, 
the $\theta$-angles are $\pi /2$, i.e. the maximally allowed values. This
state corresponds to having the spins distributed along the equator with an
angle $\pi/2$ between two successive spins. The solid angle spanned by this
configuration covers half of the unit sphere, ensuring $\Phi =\pi$. 
Thermodynamically, however, it seems to be favorable for the system to 
phase-separate into regions consisting of a hole-free antiferromagnet and a 
hole-rich coplanar structure, respectively.
If we consider small dopings ($x\ll 1$), the $\theta$-angle of the optimal 
state (coplanar) is successively reduced to zero as $x\rightarrow 0$. 
Since the maximum amplitude of the hopping for a staggered flux-state is
$|\tau|=\frac{t}{\sqrt{2}}$ as determined by $\theta\leq\frac{\pi}{2}$, 
it clearly has a disadvantage compared to the coplanar states, supporting 
$|\tau|=t$. The effect of this is that the coplanar state will always be 
favorable at fillings where the kinetic energy is dominant. However, as the
filling is further increased, the Heisenberg energy becomes more important and
it becomes favorable to reduce $|\tau|$ in order to improve the Heisenberg
energy. At this point, if this occurs at a suitable filling, the flux-state
can yield equally good Heisenberg and kinetic energies, while at the same time
providing the fermions with a flux that can lower the energy even further.
We know from the work of Hasegawa 
{\etal}~\cite{Hasegawa89} that the energy of a system of free electrons 
on a square lattice is minimized when there is exactly one flux quantum per 
particle. Since our flux-state carries a flux $\pi$ per plaquette, it will be 
most suitable close to $n=1/2$. The coplanar state on the other hand carries 
no flux, and will therefore be optimal when $n=0$ or $n=1$. This competition 
explains why the coplanar state becomes energetically favorable again when 
moving from a flux-phase towards higher fillings.

In our calculations, a slight technical point should be mentioned. When the 
system phase-separates into a coplanar part and an undoped 
antiferromagnet, the chemical potentials in the two subsystems are not equal. 
This can be seen, for instance in \Figref{free1}, where the Maxwell 
construction connects the antiferromagnetic $n_1=1$ point with a point 
$n_0\simeq 0.55$ in the coplanar phase. Since $n_1$ is an end-point of the 
free energy curve, the derivatives at $n_0$ and $n_1$ are unequal, and hence 
the chemical potential is different in the two phases.


\subsection{Results from Hartree-Fock calculations}

We now apply the Hartree-Fock scheme to the problem to understand whether or
not exchange effects can resolve the near degeneracies found in the Hartree
calculations. The spin-textures considered are those given by 
\Eqnref{spinconfig}. Before starting the full calculations, we determined
numerically which of the different order-parameters are the important ones.
Initially there were ten \cite{footop}  but numerically we find
that only four are non-zero in the spin-configurations we have examined. To
speed up the calculations we have set the other six to zero by hand.
We find that besides the 
filling ($\Delta^0$), $\Delta_{cy}^2$, and $\Delta_{cx}^3$ are important for 
the flux-phases, while $\Delta_{cx}^3$ and $\Delta_{cy}^3$ are important for 
the coplanar configurations.

Numerically we find that the $\pi$-flux-phase converges to a state where the 
non-zero order parameters are
$\Delta_{cy}^2=\Delta_{cx}^3=\Delta$, which leads to a quasiparticle spectrum
of the form $\epsilon_{\mathbf k}=-\tilde{\mu}\pm 2\tilde{t}
\sqrt{\cos^2k_x+\cos^2k_y}$, where $\tilde{\mu}$ and $\tilde{t}$ are 
renormalized values of the chemical potential and hopping 
amplitude, respectively. The momentum $\mathbf{k}$ belongs to the reduced 
Brillouin zone. At half-filling, this dispersion relation has four gapless 
Dirac-points where the energy vanishes linearly. 
These points are located at $(k_x,k_y)=(\pm\pi/2,\pm\pi/2)$.
Similarly, for the coplanar states, the mean-fields renormalize
the hopping amplitudes so that $\epsilon_{\mathbf k}=-\tilde{\mu}
\pm (\tilde{t_x}\cos k_x +\tilde{t_y}\cos k_y)$.

If we look at a typical free energy plot, such as \Figref{free1}, we find, as 
in the Hartree case, 
that for low fillings the coplanar configuration is the optimal, where at low
dopings it merges with the antiferromagnet. For small $J$'s the coplanar 
configuration clearly dominates over the flux-configuration for all fillings 
up to the point where they merge with the pure antiferromagnet. Increasing 
$J$ brings the flux-configuration energetically closer to the coplanar 
configuration. However, in contrast to the Hartree-case, it does not seem like 
the flux-state will become energetically favorable over the coplanar-states. 
Concerning phase-separation, the picture is very much the same as the one 
described above. For low temperatures and dopings smaller than roughly 0.5, 
the system favors a phase-separation into parts consisting of a hole-free
antiferromagnet and a hole-rich coplanar state with doping $x\simeq 0.5$. 

Figures~\ref{fig:free1} and \ref{fig:free2} show the free energies for the 
best coplanar- and flux-configurations plotted together with the free energy 
of the pure antiferromagnet in two different cases, $J=0.75$ and $J=2.00$. 
The energy-scale is fixed by $t=1$. The temperature is set to $T=0.1$ and the 
size of the system being considered is $16\times 16$ sites.

In \Figref{free2}, it is clearly seen how the flux-phase dominates over 
the coplanar-phase close to $n\simeq 0.5$ at the Hartree-level, but not in 
the Hartree-Fock approximation.

\begin{figure}[tbh]
\centerline{\epsfxsize=0.9\columnwidth\epsffile{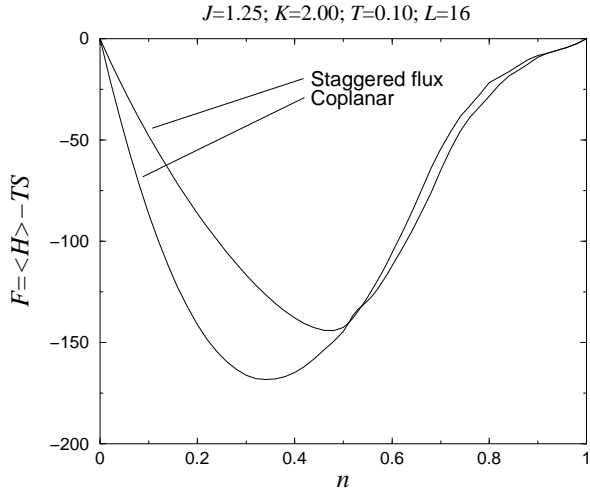}}
\caption[free3]
{\label{fig:free3}
        The Helmholtz free energy for a {\tJ} model with $J=1.25$ and
        $K=2.00$ at temperature $T=0.1$ is shown for the coplanar and
        staggered-flux spin-configurations. Note the small region about
        $n\simeq 0.53$ where the flux-configuration is energetically more
        favorable than the coplanar spin-configuration.
}
\end{figure}

We have generalized the {\tJ}-model by including a nearest-neighbor Coulomb
repulsion through a term $V\sum_{\br,\brp}n_{\br}n_{\brp}$. In our effective
model this corresponds to redefining $K_{\br\brp}\longmapsto K_{\br\brp}+V$.
When including this term the order-parameter corresponding to a 
charge-density-wave, $\Delta^1=\sum_{{\mathbf k}\in {\rm BZ'}}\langle
\Psi_{\mathbf k}^{\dagger}\sigma_1\Psi_{\mathbf k}\rangle$, becomes important.
As it turns out, a positive value of $V$ can favor the flux-phase as is seen
in \Figref{free3} where we have shown the free energy versus filling for a 
system described by $J=1.25$, $K=4V/J=2.00$, and temperature $T=0.1$. As can 
be seen from this figure, there appears a narrow region around $n\simeq 0.53$
where the flux phase has the best energetics.
\begin{figure}[tbh]
\centerline{\epsfxsize=0.9\columnwidth\epsffile{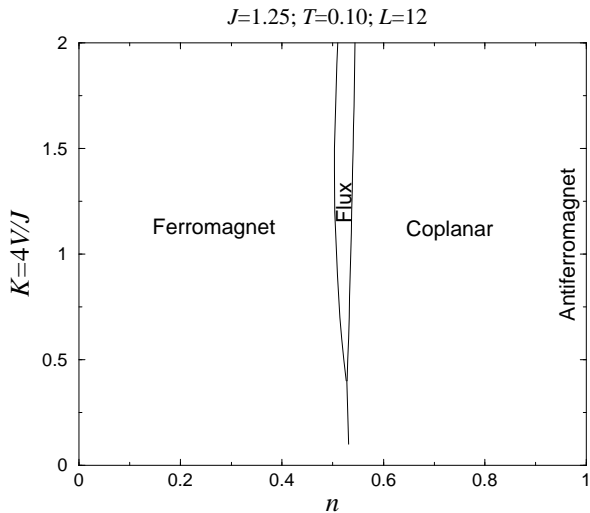}}
\caption[phaseK]
{\label{fig:phaseK}
This figure shows the state which minimizes the free energy depending
on doping and the coupling constant $K$. We have chosen $J=1.25$,
temperature $T=0.10$, and the linear dimension of the system is $L=12$. The
data is based on Hartree-Fock calculations. As can be seen, a narrow region
where the flux-configuration is the best uniform state appears around
$n\simeq 0.53$. The coplanar phase continuously merges with the
antiferromagnet as $n\rightarrow 1$.}
\end{figure}

Finally, in \Figref{phaseK}, we show a phase-diagram as a function of
filling and coupling $K$. In this data we have fixed the exchange coupling 
$J=1.25$ and temperature $T=0.1$. The figure shows how a narrow region of a 
staggered-flux phase occurs close to $n\simeq 0.5$.

Summarizing our numerical results, we find that a staggered flux-phase can
be energetically favorable compared to a coplanar spiral state. However, in
the region of doping where this happens, the system generically seems unstable
against phase-separation. Even if this indicates that the flux-phase
does not occur as a thermodynamically stable phase in our effective model, it
has good energetics and it would be interesting to investigate the effects of
spin-fluctuations in this picture.

In the next section we discuss stripes within this framework.
We know that stripes are antiphase boundaries
between antiferromagnetically ordered regions. One way to model this is to
have a spin-order twisting as the boundary is crossed. Since we also know that
the stripes are doped, it is tempting to think that the twisting is such that
a fictitious flux is generated. The picture is also appealing since it makes
use of the instability of the doped antiferromagnet towards phase-separation.
We will return to this topic in \Secref{stripes}


\subsection{Circulating currents}

An important issue to address is whether or not the flux-states are 
accompanied by circulating currents in the system. To answer this question, we
consider the current operator at site $\mathbf{r}$ in the $\delta$-direction, 
$j_{\delta}\mathbf{(r)}$, which can be identified from
charge conservation together with the Heisenberg equation of motion,
\begin{equation}
\sum_{\delta =\widehat{x},\widehat{y}}
	\left[ j_{\delta}({\mathbf r})-j_{\delta}({\mathbf r}-\delta)\right]
=-\frac{\partial\rho (\mathbf{r})}{\partial t}
=-\frac{i}{\hbar}[H_{\rm eff},\rho (\mathbf{r})] .
\end{equation}
The result is a current operator taking the form
\begin{equation}
j_{\delta}({\mathbf r})=-\frac{i}{\hbar}
	\left( \tau_{\br ,\br +\delta}c_{\br}^{\dagger}
	c_{\br +\delta}-{\rm h.c.}\right) .
\end{equation}

\begin{figure}[tbh]
\centerline{\epsfxsize=0.5\columnwidth\epsffile{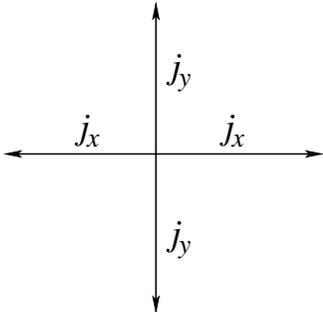}}
\caption[kirchoff]
{\label{fig:kirchoff}
        A site in the even sublattice is shown. Current conservation
        takes the form $2(j_x+j_y)=0$, which we also observe in our numerics.
}
\end{figure}

We decompose the current into uniform- and staggered parts as $j_{\delta}(\br)
=j_{\delta}^u(\br)+(-1)^{\br}j_{\delta}^s(\br)$. Using the mean-field 
decomposition we find that the expectation value of the uniform currents 
vanish.  The uniform currents are proportional to some of the 
order-parameters that have been left out of the discussion. We know that our
Hamiltonian is invariant under a lattice translation followed by a 
time-reversal operation, but this composite operation reverses the direction 
of the uniform currents which hence must vanish in a thermodynamic expectation 
value. The staggered currents take the form
\begin{eqnarray}
\langle j_{x}^s(\mathbf{r})\rangle & = & \frac{2t}{N\hbar}
		\sin\left(\frac{\theta_x}{2}\right)\Delta_{cx}^2 \cr
\langle j_{y}^s(\mathbf{r})\rangle & = & \frac{2t}{N\hbar}
		\sin\left(\frac{\theta_y}{2}\right)
		\left(  \Delta_{cy}^2\cos\frac{\Phi}{2}-
			\Delta_{cy}^3\sin\frac{\Phi}{2}\right) .
\end{eqnarray}
The currents are gauge invariant, and the formal lack of symmetry 
between the currents in the $x$- and $y$-directions is due to gauge choice.
The order-parameters also are gauge dependent, and this restores the symmetry 
between currents in the two directions, see \Figref{kirchoff}.

According to the previous subsection the non-zero order-parameters 
in the $\pi$-flux state are  $\Delta_{cx}^{3}$ 
and $\Delta_{cy}^2$, and we conclude that there are no 
circulating currents in the $\pi$-flux state. Obviously, the flux-free case
is also free of currents, as can be seen from the equations above, in this
case it is $\Delta_{cy}^3$ rather than $\Delta_{cy}^2$ which is non-zero.
Noting that in the case $\Phi =\pm\pi$ the Hamiltonian is time-reversal 
invariant so the the one-particle states possibly carrying current must 
be degenerate in energy.  In a thermodynamic ensemble, all states of the same energy are weighted 
equally, and the physical currents in a system cancel and there will
be no staggered currents in the system. For any flux not being an integer 
multiple of $\pi$, this symmetry is lost and circulating currents appear.

If we calculate the above expectation values in a state having a flux 
$\Phi\neq n\pi$, we find circulating currents in the system. 

A proper treatment of this problem should consist of a gauge invariant coupling
to a real electromagnetic
flux in addition to the spin-generated flux~\cite{Harris89}. Let $\Phi_{{\rm
mag}}$ denote the real electromagnetic flux through a plaquette, induced by
circulating currents. The energy cost of creating the magnetic field should be
added through a term
\begin{equation}
E_{\rm mag}=\frac{K}{2}\sum_{{\rm plaquettes}}\Phi_{\rm mag}^2.
\end{equation}
The constant $K$ is given by $K=\frac{dh^2}{\mu_0\mu_re^2a^2}$, where $a$ is 
the two dimensional lattice constant, $d$ the distance between the copper
oxide planes, and $\mu_r$ the relative permeability. As was discussed in 
\Secref{adiabatic},
this constant is huge compared to the typical electronic energies.
The total energy as a function of the electromagnetic flux is then written as
\begin{equation}
E(\Phi_{{\rm mag}})=E_{t-J}(\Phi_{\rm mag})+E_{\rm mag}(\Phi_{\rm mag}),
\end{equation}
where $E_{t-J}(\Phi_{\rm mag})$ denotes the energy of the {\tJ} model when 
there is an extra flux $\Phi_{\rm mag}$ in addition to the spin-generated flux.
Minimizing the energy with respect to the electromagnetic flux leads 
to an equation of the form $g(\Phi_{\rm mag})+K\Phi_{{\rm mag}}=0$,
where $g(\Phi_{\rm mag})=E'_{t-J}(\Phi_{\rm mag})$.
As we have pointed out, the magnetic energy-scale $K$ is much 
larger than the electronic energy which is of the order $\max (t,J)$. As a 
consequence of this, the magnetic flux will be suppressed and it is 
reasonable to put $\Phi_{\rm mag}=0$ when we solve for the eigenstates of the
system. For this to be a self-consistent solution there must be no currents in
the system due to the spin-generated flux. This is true for the cases $\Phi=0$
or $\pi$ which we have focused on.



\section{Striped structures}
\label{sec:stripes}

Striped structures forming
antiphase domain walls between undoped antiferromagnetic regions have been
experimentally observed in the doped high-temperature 
superconductors~\cite{Tranquada95}. We would like to understand if a striped
phase can be explored using the effective {\tJ} Hamiltonian, \Eqnref{tjeff}. 
There are
several facts that make this an appealing approach. We have already seen
in the previous section that for low dopings, the uniform (spiral) states
are unstable against phase-separation. Using our spin-polarized 
approach we can create a smooth antiphase boundary, successively changing the
order-parameter from $+\widehat{z}$ to $-\widehat{z}$. And, since all the
holes of the stripe are located in the domain wall it could be favorable for
the system to generate a flux in this region. This can be accomplished using
a spin-texture, which simultaneously generates the antiphase boundary. 
Furthermore, the experimentally observed value of the doping of stripes in
La$_{2-x}$Sr$_x$CuO$_4$ is $0.5$ holes per unit length of the 
stripe~\cite{Tranquada95}. This is close to 
the region where we have seen that a $\pi$-flux may be 
favorable. (See \Secref{fermionswithflux}.)
Inspired by these nice properties we have undertaken an investigation of 
striped phases within our approach. Our
main ambition has been to gain an understanding of what such a striped phase
would look like, and, in particular, whether a fictitious flux is exploited
or not in our model. The technique we use is a self-consistent Hartree 
calculation.

\begin{figure}[tbh]
\centerline{\epsfxsize=\columnwidth\epsffile{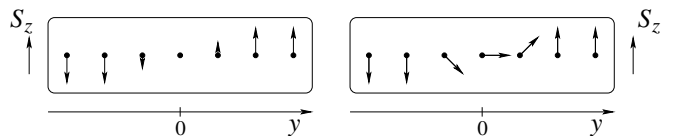}}
\vspace{2mm}
\caption[antiphasestripe]
{\label{fig:antiphasestripe}
        Two possible spin-orders making up an antiphase boundary between two
        N{\' e}el-ordered regions. The $y$ coordinate is orthogonal to the
        direction of the stripe, and $y=0$ corresponds to the center of the
        stripe. The left part of the figure shows the
        antiferromagnetic order-parameter passing through zero, while the right
        part shows our scenario with the order-parameter tilting as one passes
        through the stripe.
}
\end{figure}

Let us consider a system with an antiphase boundary along the $x$-axis, and
with the antiferromagnetic order being $+\widehat{z}$ at $y=\infty$ and
$-\widehat{z}$ at $y=-\infty$.
Two possible scenarios for an antiphase spin-order in a stripe comes to mind. 
First, the amplitude of the antiferromagnetic order-parameter may simply
decrease, passing through zero and becoming negative as one passes through the
stripe. This scenario preserves the rotational symmetry about the spin 
$z$-axis.
The second possibility is that the N{\'e}el order-parameter starts to tilt as 
one approaches the stripe along the $y$-axis, lying within the spin 
$x$-$y$-plane at the center of the stripe
and then rotating towards the positive $z$-axis as one goes to $y=\infty$.
Introducing holes in the domain wall, the amplitude of the 
spin-order will decrease as it depends on the particle density through 
${\mathbf S}_{\mathbf r}=\frac{1}{2}n_{\mathbf r}\widehat{\Omega}_{\mathbf r}$,
but it will not vanish. An illustration of these two scenarios is 
shown in \Figref{antiphasestripe}. From an experimental point of view, 
Tranquada and coworkers argue~\cite{Tranquada97,Tranquada96} that there are
indications speaking in favor of the first scenario, but these indications
are not conclusive.

\begin{figure}[tbh]
\centerline{\epsfxsize=\columnwidth\epsffile{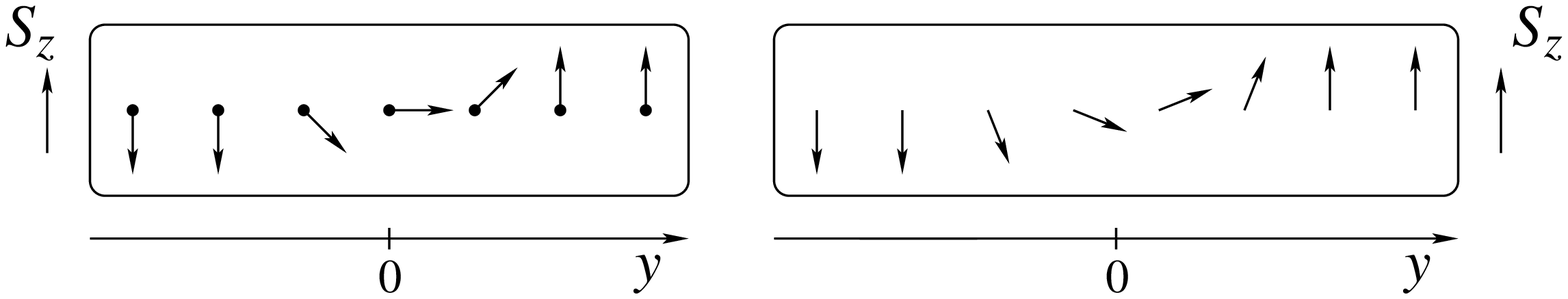}}
\vspace{2mm}
\caption[sitevsbond]
{\label{fig:sitevsbond}
        To the left is shown a site-centered stripe, in which there is a site
        at which the spin-order is in the $x$-$y$-plane. In the bond-centered
        case (right), the antiferromagnetic
        spin-order resides in the $x$-$y$-plane at an imagined
        point between two sites.
}
\end{figure}

Four different stripe geometries with preserved uniaxial symmetry come to 
mind. First, stripes can go along either the $(10)$- or $(11)$-directions and
we can choose either site- or bond-centered stripes, all in all four possible
combinations. The difference between site- and bond-centered stripes is shown
in \Figref{sitevsbond}.

\begin{figure}[tbh]
\centerline{\epsfxsize=\columnwidth\epsffile{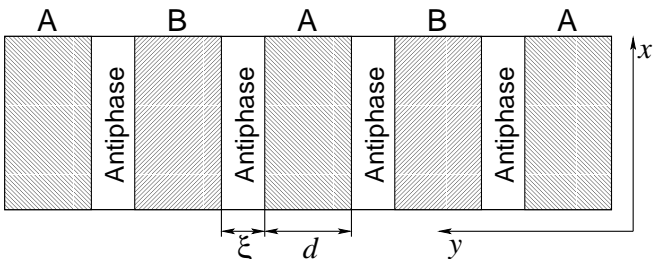}}
\vspace{2mm}
\caption[stripedphase]
{\label{fig:stripedphase}
        Our model of the striped phase is shown. $A$ and $B$ denote the two
        antiferromagnetic phases (order-parameter $\pm1$) and in between each
        $A$ and $B$ pair is an antiphase domain wall reversing the sign of the
        antiferromagnetic order-parameter. The figure also introduces $\xi$ as
        the width of the antiphase boundary, and $d$ as the size of the
        antiferromagnetic layers between the stripes.}
\end{figure}

Our aim is to find the distribution of holes and spin-texture that minimizes 
the energy of an antiferromagnet, when we assume the holes are arranged in
stripes. The configurations we consider have the structure 
shown in \Figref{stripedphase}, consisting of a repeated structure of 
N{\'e}el-regions and antiphase domain walls. The order-parameter of the 
antiferromagnetic regions change sign every time a domain wall is passed.

We will assume that all the holes are located near the
domain walls. Following the notation in 
\Figref{stripedphase}, we will denote the width of a single antiphase boundary
by $\xi$, and the width of an antiferromagnetic region by $d$. The average 
particle density in the domain wall will be denoted $n_d$, and we will also 
use the number of holes per unit length (along $x$) of the domain wall, 
$\delta=(1-n_d)\xi$. Assuming an average filling $n$ of the system, we have 
$(\xi +d)n=\xi n_d+d$.

Before writing down the total energy of this configuration, we make the 
following observations. We note that the system in \Figref{stripedphase} is
build up from units (antiferromagnets and domain walls) and we would like to 
write the total energy in terms of the energies of the individual units. To do
this we define the energy per site of an antiferromagnetic unit, $E_{AF}$, as 
the energy per site of the antiferromagnet with periodic boundary conditions 
in the $x$- and $y$-directions. The energy per site
of the domain wall as a function of the number of holes per unit length
of the stripe, $E_d(\delta)$, is similarly defined by putting periodic boundary
conditions on the domain wall. In this case it is important that the edge of
the domain wall is N{\'e}el-ordered and undoped. If this condition is not 
fulfilled, there will be surface energies associated with the gluing of a 
domain wall to an antiferromagnetic region. It is easy to see that 
minimizing the total energy of the system is 
equivalent to minimize the energy per introduced hole in the domain wall.



Our approach will therefore be to consider a single domain wall and minimize
the energy per introduced hole with respect to the parameters describing the
domain wall. These parameters contain spin-texture related parameters (which
will be introduced shortly), the number of holes per unit length of the stripe
($\delta$), and finally we have the four options for the stripe geometry;
site/bond-centered and direction (10)/(11).

Note that the
definition of $\xi$ is somewhat arbitrary in the sense that there in practice
may be a smooth crossover from the domain wall to the N{\'e}el-ordered 
region. For this reason, $\delta$ is a better measure of the stripe-doping
than $n_d$ which depends on $\xi$. Knowing $\delta$ and $\xi$, the 
stripe periodicity is
\begin{equation}
\label{eq:stripesep}
l=2(\xi +d)=\frac{2\delta}{x},
\end{equation}
where we have introduced the average doping of the system, $x=1-n$. Thus we
find that for low dopings, the separation between the stripes scales as 
$x^{-1}$. This description is valid as long as the stripes remain separated
so that we can neglect stripe-stripe interactions. This condition is fulfilled
as long as $x\ll\frac{\delta}{\xi}$. We note
that as we change the overall doping of the system, the structure of the 
isolated stripes remains, at least as long as $d\gg 1$. Thus the wavevector
describing the spin order is $2\pi/l$, and the wavevector of the charge order
is twice that, i.e. $4\pi/l$.


\subsection{Stripes in the $(10)$-direction}

Let us start with a description of how a single $(10)$-stripe is modeled.
Although we are interested in infinite domains, in the numerical simulations
we are forced to work with finite stripe width, which we denote by the integer
$w$.
We will assume the local spin orientation of the stripe to described by a 
unit-vector field $\widehat{\Omega}_{\mathbf r}$ as
\begin{equation}
\label{eq:stripeconfig}
\widehat{\Omega}_{\mathbf r} = (-1)^{\mathbf r}(\sin\theta_{\mathbf r}
	\cos\phi_{\mathbf r},\sin\theta_{\mathbf r}\sin\phi_{\mathbf r},
	\cos\theta_{\mathbf r}) ,
\end{equation}
where the spherical
angles $\theta_{\mathbf r}$ and $\phi_{\mathbf r}$ are functions of 
position. The spatial dependence of these angles is assumed to take the 
form
\begin{eqnarray}
\theta_{\br}	&=&	\arccos\left[\tanh\frac{w+1-2y}{2\xi}\right] \cr
\phi_{\br}	&=&	q_x x .
\label{eq:stripespinorder}
\end{eqnarray}
The construction can be thought of as a cylindrical projection (also known as
Mercator projection) of the lattice into the spin sphere by associating
latitude and longitude with the $x$- and $y$-coordinates, respectively.

\begin{figure}[tbh]
\centerline{\epsfxsize=0.9\columnwidth\epsffile{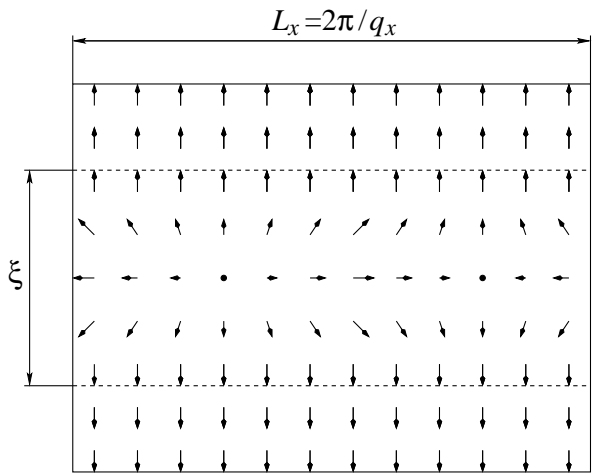}}
\vspace{3mm}
\caption[stripe]
{\label{fig:stripe}
        A sketch of the N{\'e}el order-parameter in the neighborhood
        of a stripe constituting an antiphase boundary between two regions with
        perfect N{\' e}el order. All spins are unit vectors. In the figure we
        have also indicated the length scale $\xi$. $L_x$, the period along
        the $x$-direction, is related to $q_x$
        through $q_x=2\pi /L_x$. This particular spin-configuration is clearly
        site-centered.
}
\end{figure}

We are interested in the limit $w\rightarrow\infty$. This limit has two 
discrete values, depending on whether $w$ is even or odd. For odd values of 
$w$ the spin-order of
\Eqnref{stripespinorder} is such that the spins at $y=(w+1)/2$ are lying in 
the $x$-$y$-plane, and hence the stripe is site-centered. If we instead 
consider an even integer $w$, the center of the stripe is located in between 
two rows and the stripe is bond-centered.

In addition to the discrete choice of bond- or site-centered stripe,
there are two free continuous parameters in \Eqnref{stripespinorder}; 
$\xi$, which determines the characteristic width of the spin-domain wall, 
and $q_x$, which is the pitch of the rotation of the spin about the $z$-axis
along the length of the stripe. \Figref{stripe} illustrates this construction.

Making an apparently trivial point, which we return to in the next section,
we note that there is no choice for
$\xi$ and $q_x$ which unwraps the antiferromagnet into a ferromagnet; taking
$q_x=\pi$ will create a ferromagnetic channel in the center of the domain wall,
running along the stripe, but will not affect the $z$ component of the
N{\'e}el order-parameter.

When writing down the Hamiltonian of this system we need to fix a gauge. If the
spin-configuration around a square plaquette in terms of spherical angles is
given by: $(\theta_1,\phi_1)$, $(\pi -\theta_2,\phi_1+\pi)$,
$(\theta_2,\phi_2)$, and $(\pi -\theta_1,\phi_2+\pi)$,
the product of the hopping elements around the plaquette is
\begin{eqnarray}
\tau_{01}&\tau_{12}&\tau_{23}\tau_{30}=\cr &&-\sin\theta_1\sin\theta_2\sin^2
	\frac{\theta_1-\theta_2}{2}\sin^2\frac{\phi_1-\phi_2}{2}\leq 0,
\end{eqnarray}
showing that the flux through a plaquette is exactly equal to $\pi$. The
texture is therefore quite natural for the following reasons: It
is periodic and has a uniform flux per square. The flux $\pi$
per plaquette favors heavily doped regions near the center of the stripe,
while the effective hopping in the N{\'e}el regions, where the system is
undoped, is vanishing. With flux $\pm\pi$ through the plaquettes there is no
broken time-reversal symmetry, and there are no circulating currents or 
induced local magnetic fields.

In order to perform our calculations, we use a mixed representation using 
momentum space in the $x$-direction and real space in the $y$-direction
due to the translational invariance of the system along the $x$-direction.
The length of our system along the stripe will be denoted by $L$. Furthermore 
we assume that the number density at a site $\br$ only depends on the 
$y$-coordinate, i.e. $n(\br )=n(y)$. We will use
the following Hartree-decomposition of the interaction term in 
\Eqnref{tjeff}
\begin{equation}
n_{\br}n_{\brp}\simeq n(y)n_{\brp}+n_{\br}n(y')-n(y)n(y').
\end{equation}
The effective Hamiltonian can then be written as
\end{multicols}
\widetext
\begin{eqnarray}
H_{\rm bulk}&=&-2t\sum_{y=1}^w(-1)^y\sum_k\cos (k)\tau_x(y)n_{k,y}
	-t\sum_{y=1}^{w-1}\tau_y(y)\sum_k
	\left[c_{k,y+1}^{\dagger}c_{k,y}+\hc\right] \cr
&& 	+\frac{J}{4}\sum_{y=1}^w(h_x(y)-1)\sum_k
	\left[ 2n_{k,y}n(y)-n(y)^2\right] \cr
&&	+\frac{J}{4}\sum_{y=1}^{w-1}(h_y(y)-1)\sum_k[ n_{k,y}n(y+1)
	+n_{k,y+1}n(y)-n(y)n(y+1)] ,
\label{eq:stripe10ham}
\end{eqnarray}
\begin{multicols}{2}
\narrowtext
\noindent
where we have introduced the coordinate-dependent coupling constants
\begin{eqnarray}
h_{x}(y)	&=&	\widehat{\Omega}_{{\mathbf r}}\cdot
			\widehat{\Omega}_{{\mathbf r}+\widehat{x}} \cr
\tau_{x}(y) 	&=&	\left(\frac{1+h_x(y)}{2}\right)^{\frac{1}{2}}, 
\end{eqnarray}
and the analogous relations for $h_y(y)$ and $\tau_{y}(y)$. The factor $(-1)^y$
in the hopping term of \Eqnref{stripe10ham} comes from the flux $\pm\pi$ 
through each plaquette. In addition to the
terms in \Eqnref{stripe10ham}, we have also added a local
potential at the upper and lower edges of the system to simulate the effect
of the adjoining antiferromagnetic region. Without this potential, which
has the form
\begin{equation}
\label{eq:10edgepot}
H_{\rm b.c.} = -\frac{J}{2}\sum_k\left[n_{k,1}+n_{k,w}\right],
\end{equation}
the system may gain energy by expelling the holes to the edges where they break
fewer antiferromagnetic links. The total Hamiltonian is then given by 
\begin{equation}
	H_{10}=H_{{\rm bulk}}+H_{{\rm b.c.}}.
\end{equation}

The numerical calculation involves solving self-consistently for a charge
profile described by $n(y)$, where
\begin{equation}
\label{eq:chargecond}
n(y)=L^{-1}\sum_k\langle n_{k,y}\rangle ,
\end{equation}
and the expectation value is with respect to the Fermi-distribution of 
quasiparticles of $H_{10}$. We use a chemical potential to 
control the overall number of particles in the system. The chemical potential
is determined during each iteration of the self-consistency equation, 
\Eqnref{chargecond}. Our calculations are performed at temperatures close to
zero, $T=0.01t$, and the quantity we focus on is the free energy as a function
of the number of particles and the parameters describing the stripe, 
$F(T,N)=\langle H_{\rm 10}\rangle -TS$.


\subsection{Stripes in the $(11)$-direction}

\begin{figure}[tbh]
\centerline{\epsfxsize=0.8\columnwidth\epsffile{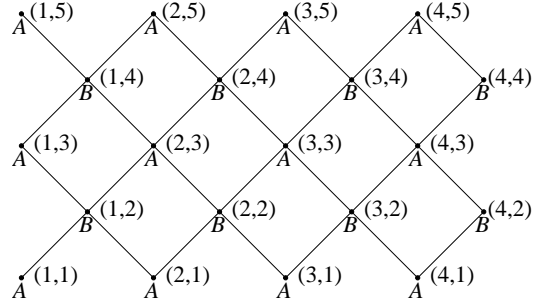}}
\vspace{2mm}
\caption[11stripe]
{\label{fig:11stripe}
        The geometry used for the description of $(11)$-stripes is shown.
        The original square lattice has been tilted 45 degrees, and we have
        labeled each point with a coordinate $(x,y)$. These coordinates are
        not be confused with the coordinates in the original, untilted square
        lattice. Each lattice point has also been marked with the sublattice,
        $A$ or $B$, to which it belongs. The domain wall is directed along
        the $x$-axis in this coordinate system, and coupling constants will
        depend on the $y$-coordinate only.
}
\end{figure}

The analysis of the $(11)$-stripes is similar, although the geometry of the
stripe introduces some complications. In \Figref{11stripe}, we have shown the
geometry used for these configurations. Due to the tilting of the lattice, we
note that every second row of constant $y$ in the lattice is shifted to the 
right by half the lattice spacing in the $x$-direction.
Moreover, if we denote the even- and odd sublattices by $A$ and 
$B$, respectively, we find that all points belonging to $A$ reside on points
having odd values of $y$, while those belonging to $B$ are assigned even values
of $y$. This is shown explicitly in \Figref{11stripe}.

The stripe is directed along the $x$-axis in the coordinate system defined in
\Figref{11stripe}, and the spin-configuration is given by
\begin{equation}
\widehat{\Omega}_{\mathbf r}=(-1)^y(\sin\theta_{\mathbf r}\cos\phi_{\mathbf r},
	\sin\theta_{\mathbf r}\sin\phi_{\mathbf r},\cos\theta_{\mathbf r}).
\end{equation}
We note that $\theta_{\mathbf r}=0$ and $\theta_{\mathbf r}=\pi$ correspond
to the two N{\'e}el-states, and hence we can (as in the $(10)$-case) 
interpolate between the two by continuously changing $\theta_{\mathbf r}$
across the domain wall. To be explicit, we will use the following 
parametrization of the spherical angles:
\begin{eqnarray}
\theta_{\mathbf r}	&=&	
		\arccos\left[ \tanh\frac{w+1-2y}{2\xi}\right]\cr
\phi_{\mathbf r}	&=& \left\{
	\begin{array}{ll}
		q_x x	&	y\hbox{ odd} \\
		q_x(x+1/2) &	y\hbox{ even} \\
	\end{array}
	\right. .
\label{eq:11config}
\end{eqnarray}
Note that we have shifted $x$ by $1/2$ for even $y$-values to account for the
shift of lattice points in the $x$-direction as discussed 
above. It is also important to stress that, contrary to the 
$(10)$-case, it is possible to recover a ferromagnetic configuration by a
suitable choice of parameters, namely making $\xi$ large and taking 
$q_x=2\pi$.
This corresponds to $\theta_{\bf r}=\pi/2$ and $q_x=2\pi$ for odd $y$'s and
$q_x=\pi$ for even $y's$, respectively. The effect of the rotation due to $q_x$
is therefore to rotate all spins belonging to sublattice $B$ by $\pi$ about
the spin $z$-axis. The result is a ferromagnetic configuration, where all spins
point along the positive $x$-axis.
This difference between the $(10)$- and $(11)$-stripes
reflects the fact that the local field along either side of a $(11)$-stripe
is ferromagnetic whereas in the $(10)$ case it is 
antiferromagnetic~\cite{Japaridze}.

\begin{figure}[tbh]
\centerline{\epsfxsize=0.8\columnwidth\epsffile{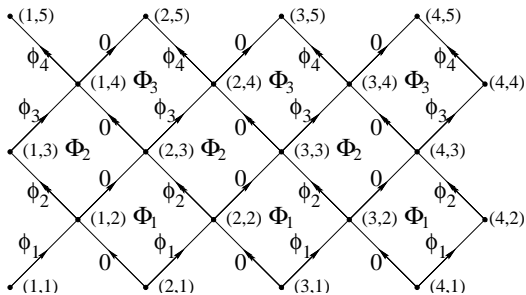}}
\vspace{2mm}
\caption[11flux]
{\label{fig:11flux}
        The lattice of the $(11)$ domain wall is shown. The structure of the
        flux pattern is given by the $\Phi_y$'s. In particular we find that
        the flux is uniform along the $x$-axis. We have fixed a gauge by
        defining the phases associated with the links in the lattice. The
        gauge is chosen such that all non-zero phases are on links connecting
        sites with the same $x$-coordinate.
}
\end{figure}

The next issue we will address is the properties of the fictitious fluxes 
generated by a certain spin-configuration in the domain wall. \Figref{11flux}
shows the flux-pattern which is generated from the spin-configuration in
\Eqnref{11config}. 

We find that the flux is uniform along the $x$-direction, while it depends on
the $y$-coordinate of the plaquette to which it belongs. \Figref{11flux} also
fixes a gauge, defined by the phases $\phi_y$. As we can see from the figure,
there is one phase more than there are gauge invariant fluxes. Therefore, we
can choose a gauge in which $\phi_1=0$. Next, we consider the flux $\Phi_1$
which is determined by $\Phi_1=\phi_1+\phi_2$. More generally, we find
$\Phi_y=(-1)^{y+1}(\phi_y+\phi_{y+1})$. In this way we can solve for the 
phases $\phi_y$ in terms of the fluxes, finding
\begin{equation}
\label{eq:11gaugechoice}
\phi_{y+1}=(-1)^{y}\sum_{y'=1}^{y}\Phi_{y'}.
\end{equation}
Numerically, given the spin-configuration of the domain wall we 
calculate the fictitious fluxes, $\Phi_y$, using \Eqnref{flux} and then use
\Eqnref{11gaugechoice} to find the appropriate phases that enter the 
Hamiltonian.

We define the amplitudes of the hopping and Heisenberg interactions as
\begin{eqnarray}
h_y(y)	&=&	\widehat{\Omega}_{x,y}\cdot
		\widehat{\Omega}_{x,y+1} \cr
h_d(y)	&=&	\widehat{\Omega}_{x,y}\cdot
		\widehat{\Omega}_{x+(-1)^y,y+1}\cr
\tau_y(y)&=&	\left(\frac{1+h_y(y)}{2}\right)^{\frac{1}{2}} \cr
\tau_d(y)&=&	\left(\frac{1+h_d(y)}{2}\right)^{\frac{1}{2}},
\end{eqnarray}
which allows us to write the bulk part of the Hamiltonian as
\end{multicols}
\widetext
\begin{eqnarray}
\label{eq:stripe11ham}
H_{\rm bulk} &=& \sum_{y=1}^{w-1}\sum_k\Bigl[ -t\left(
		\tau_y(y)e^{i\phi (y)}+\tau_d(y)e^{i(-1)^yk}\right)
		c_{k,y+1}^{\dagger}c_{k,y}+\hc\Bigr]\cr
	&&	+\sum_{y=1}^{w-1}\frac{J}{4}\left( h_y(y)+h_d(y)-2\right)
		\sum_k\left[ n_{k,y}n(y+1)+n_{k,y+1}n(y)-n(y)n(y+1)\right].
\end{eqnarray}
\begin{multicols}{2}
\narrowtext
\noindent
For the same reasons as in the $(10)$-case, we will add a local potential to
the vertical boundaries (note that each boundary site connects to two sites
in the environment)
\begin{equation}
H_{\rm b.c.}=-J\sum_k\left[n_{k,1}+n_{k,w}\right].
\end{equation}
The total Hamiltonian, $H_{11}$, is the sum of the bulk- and 
boundary-contributions, i.e.
\begin{equation}
H_{11}=H_{\rm bulk}+H_{\rm b.c.}.
\end{equation}
The numerical procedures are completely equivalent to those used in the 
$(10)$-case.  


\subsection{The optimal stripe}
\label{sec:optstripe}

According to our model, the physically relevant stripe configuration is that 
which minimizes the domain wall energy per introduced hole. We will use the 
undoped antiferromagnet as an energy reference state, this being the optimal 
state at zero doping.
As we dope holes into the system, the total energy of the domain wall will 
be a discrete function of the geometry (site- or 
bond-centered and direction (10) or (11)) and a continuous 
function of $N_h$, $q_x$, and $\xi$. Note
that since we work with a chemical potential, the number of holes in the
domain wall, $N_h$, is not restricted to be an integer.

We 
label this energy $E(N_h,q_x,\xi)$. However, the physically interesting 
quantity is the number of holes per unit stripe length, $\delta =N_h/L$. We
define the domain wall energy per hole according to
\begin{equation}
E_h(\delta ,q_x,\xi )	=\frac{E(\delta L,q_x,\xi)-E_{\rm AF}}
			 {\delta L} .
\label{eq:eperhole}
\end{equation}
As we argued in the beginning of this section,
to find the optimal stripe-configuration we have to minimize this function
with respect to $\delta$, $q_x$, and $\xi$ for bond- and site-centered $(10)$-
and $(11)$-stripes.

\begin{figure}[tbh]
\centerline{\epsfxsize=0.9\columnwidth\epsffile{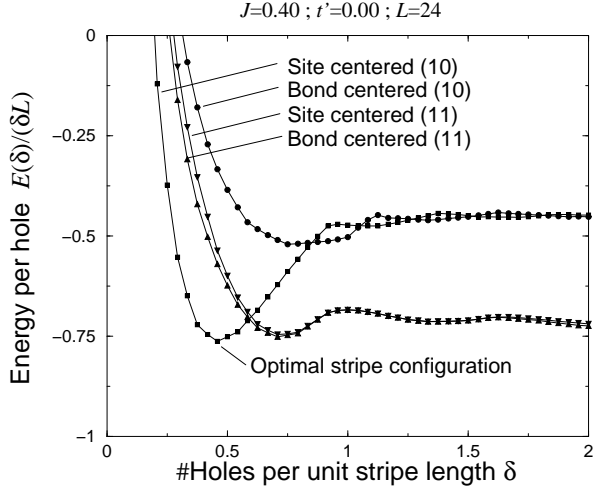}}
\caption[EPerHoleJ040L24]
{\label{fig:EPerHoleJ040L24}
        The energy per hole is shown as a function of the number of holes per
        unit stripe length, $\delta$. An investigation of the curves shows that
        the minimum occurs for the site-centered $(10)$-stripe at
        $\delta\simeq 0.46$.
}
\end{figure}

Turning to our numerical results, we have initially considered a system with
a Heisenberg coupling $J=0.40$, where the energy scale is fixed by $t=1$. This
value was chosen because it corresponds to a value of the exchange coupling 
constant which has been used by others to model the high-temperature 
superconductors.
In \Figref{EPerHoleJ040L24} we show the optimal energy per hole as a function
of $\delta$ for the four stripe geometries, i.e. we have plotted
$E_h(\delta)=\min_{q_x,\xi}E_h(\delta ,q_x,\xi)$.

\begin{figure}[tbh]
\centerline{\epsfxsize=\columnwidth\epsffile{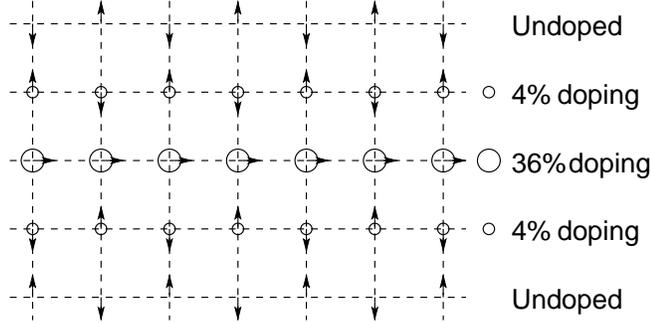}}
\vspace{2mm}
\caption[10profile]
{\label{fig:10profile}
        The structure of the optimal $(10)$-stripe for $J=0.40$ is shown.
        Arrows indicate the polarization of the spins, and the radius of the
        circles the amount of hole-doping. Small circles correspond to
        $4~\%$ hole-doping, while large circles correspond to $36~\%$
        hole-doping. Undoped regions lack circles.
}
\end{figure}

From \Figref{EPerHoleJ040L24} we can read off the optimal stripe 
configuration, which will make up the domain walls in the striped phase. As is
indicated in this figure, the optimal domain wall is a 
site-centered stripe along the $(10)$-direction having $\delta\simeq 0.46$ 
holes per unit length of the stripe. This agrees with results from DMRG 
calculations by White and collaborators~\cite{White98}, who find stripes with 
$\delta=0.5$ for $J = 0.35$. Furthermore, the experimental data indicate
that $\delta\simeq 0.5$~\cite{Tranquada95}. 
We also find that $\xi$ is very 
small for this optimal stripe, i.e. there is a sharp spin-domain wall with a 
single tilted row of spins. This row is ferromagnetically ordered as we find 
$q_x=\pi$, and the holes are tightly confined in the neighborhood of the 
domain wall. The spin- and charge-profiles are shown in \Figref{10profile}.

Since the optimal domain wall is so narrow, the product of the effective
hopping-amplitudes around any plaquette in the lattice is approximately zero.
Thus we must conclude that the system does not take advantage of the 
fictitious $\pi$-flux through the plaquettes.

\begin{figure}[tbh]
\centerline{\epsfxsize=\columnwidth\epsffile{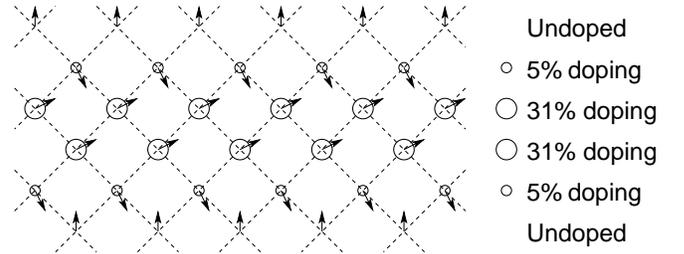}}
\vspace{2mm}
\caption[11profile]
{\label{fig:11profile}
        The structure of the optimal confined $(11)$-stripe
        for $J=0.40$ is shown.
        Arrows indicate the polarization of the spins, and the radius of the
        circles the amount of hole-doping. Small circles correspond to
        $5~\%$ hole-doping, while large circles correspond to $31~\%$
        hole-doping. Regions being approximately undoped lack circles.
}
\end{figure}

As \Figref{EPerHoleJ040L24} shows, the bond-centered $(11)$-stripe is 
energetically
very close to the optimal $(10)$-stripe described above. An illustration of
this domain wall configuration is shown in \Figref{11profile}. This 
$(11)$-stripe is characterized by $\delta\simeq 0.71$, $q_x=2\pi$, and 
$\xi\simeq 0.73$. It is important to point out that $q_x=2\pi$ is not 
equivalent to $q_x=0$ since $q_x=2\pi$ performs a $\pi$-rotation about the 
$z$-axis of one of the sublattices, as we discussed in the previous 
subsection. Furthermore, we want to emphasize that this
stripe-configuration does not induce any fictitious fluxes and consequently
there are no currents that could energetically disfavor this configuration.

Since the optimal $(10)$-stripe is very close in energy to the $(11)$-stripes
it is interesting to investigate what happens as we tune the strength of
the Heisenberg interaction, $J$. Numerically we find that slightly increasing
$J$ above $J=0.40$ 
favors the $(10)$-stripes compared to $(11)$-stripes, while decreasing $J$
favors the $(11)$-stripes. There will be a crossing point
slightly below $J=0.40$, at approximately $J\simeq 0.36$, where the 
$(11)$-stripes have lower domain wall energy than the $(10)$-stripes.

There is a technical point which is important when looking at the energy per
hole as a function of $\delta$ for the $(11)$-configurations. As we mentioned
in the previous subsection, it is possible to unwrap the $(11)$-domain walls
into ferromagnets. If we follow the $(11)$-curves in \Figref{EPerHoleJ040L24}
to larger values of $\delta$ we find that the energy successively decreases 
below what we called the optimal stripe configuration. The configurations that
correspond to these low-energy states are close to ferromagnetic and with an 
almost uniform charge-distribution. Physically this corresponds to a global 
phase-separation into an undoped N{\'e}el region and a heavily doped 
ferromagnetic region, i.e. the phase-separation discussed in 
\Secref{uniform}. Let us therefore compare the energetics of the optimal 
$(10)$-domain wall and the global phase-separation.

At the minimum of $E_h(\delta)$ in \Figref{EPerHoleJ040L24},
we can read off the energy per hole of the optimal domain wall as $E_{\rm dw}
\simeq -0.76$. This is to be compared to the energy per hole for the totally
phase-separated state $E_{\rm ps}\simeq-1.06$ for the same value of $J$.
This number was obtained from a Hartree calculation using 
\Eqnref{unihartree} on a system of the same size as
the stripe grid. This clearly shows that, within our approximation at least,
phase-separation is energetically
advantageous compared to domain wall formation. In a real system, the energy
of the phase-separated state is raised due to the Coulomb interaction between
the holes and this could make the domain wall thermodynamically stable.

The cuprates seem to favor
the formation of rather narrow stripes, not 
phase-separation. We will take the point of view that there is some mechanism,
not captured in our approach, such as a long-range Coulomb interaction, which
prevents grouping all the holes together and instead favors the formation of
stripes on some intermediate length scale. Therefore, we will only consider the
stripe-configurations which are local in nature. 


\subsection{Including next-nearest-neighbor hopping}

An unphysical feature of our simulation is that since hopping between the
antiferromagnetic spins is forbidden, the effect of second-neighbor 
(diagonal) hopping
becomes important in the N{\'e}el state. In our approximation of the {\tJ} 
model, hopping is frozen out for this state, so that electron transport will 
be dominated by any second-neighbor terms if they are non-zero. This term will
permit electrons to diffuse into the N{\'e}el state and could therefore be
expected to delocalize the holes from the stripe center.
We will extend
our model of the $(10)$-stripes by adding this hopping to the Hamiltonian 
through a term
\begin{equation}
H_{nnn}=-t'\sum_{\langle\langle {\br\br'}\rangle\rangle}\left[\tau_{\br\br'}'
c_{\br '}^{\dagger}c_{\br}+\hc\right] ,
\end{equation}
where $\langle\langle\br\br'\rangle\rangle$ denotes next-nearest-neighbor 
pairs. As before, the calculations reported below are performed with the Heisenberg
coupling $J=0.40$ and at temperature $T=0.01$ in terms of energy units set
by $t=1$. In this section, we will only consider the effect of second-neighbor
hopping on the $(10)$-stripe, since this was found to be the optimal 
stripe-configuration for $J=0.40$.

It has been argued in the literature that the sign of $t'$  in the 
high-$T_c$ cuprates depends on whether the system is hole-doped
or electron-doped~\cite{Tohyama94,Gooding94}. A hole-doped system corresponds
to $t'<0$, while an electron-doped system has $t'>0$. In an antiferromagnet,
the presence of a second-neighbor hopping is important since it allows for
holes moving through the sublattices without disrupting the antiferromagnetic
order. Typically, the value of $t'$ used in the literature for describing a 
hole-doped antiferromagnet is $t'\simeq -0.3$.

If we consider the limit in which the nearest-neighbor hopping is completely
frozen out, and there is only second-neighbor hopping, i.e.
\begin{eqnarray}
H_{t'-J}&=&-t'\sum_{\langle\langle\br\brp\rangle\rangle}\left[ 
	\tau_{\br\brp}'c_{\brp}^{\dagger}
	c_{\br}+\hc\right]\cr &&+\frac{J}{4}\sum_{\langle\br\brp\rangle}
	\left(\widehat{\Omega}_{\br}\cdot\widehat{\Omega}_{\brp}-1\right)n_{\br}n_{\brp},
\end{eqnarray}
we note that the transformation $c_{\br}\longmapsto (-1)^xc_{\br}$ leaves the
Heisenberg term unchanged, while the second-neighbor term changes sign. This
shows that in this limit the asymmetry between $\pm t'$ vanishes. Hence, the
sign of $t'$ is irrelevant in the N{\'e}el-ordered regions, and it is only
in the region where the spin-twist occurs that $t$ and $t'$ are simultaneously
present and accordingly, the sign of $t'$ is important.

\begin{figure}[tbh]
\centerline{\epsfxsize=0.9\columnwidth\epsffile{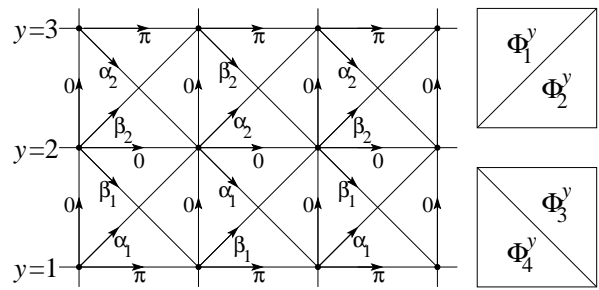}}
\vspace{2mm}
\caption[nnngauge]
{\label{fig:nnngauge}
        Our gauge choice for the next-nearest-neighbor hopping. The gauge
        choice on the horizontal- and vertical links is the same as the one
        used previously for the $(10)$-domain walls.
        The right inset defines the fluxes $\Phi_k^y$ through
        the four sub-plaquettes of a square plaquette.
}
\end{figure}

Introducing next-nearest-neighbor hoppings as in \Secref{fluxproperties} 
allows for new closed particle-orbits in the lattice,
and hence also for new gauge invariant fluxes. There are four of these fluxes
and they are defined in the right inset of \Figref{nnngauge}.

Investigating the spin-structure of Eqs.~(\ref{eq:stripeconfig}) and 
(\ref{eq:stripespinorder}), we find that the four fluxes 
$\{\Phi_k^y\}_{k=1}^4$ associated with row $y$ 
are determined by a single parameter $\Psi_y$ according to
\begin{equation}
\label{eq:subfluxes}
\Phi_1^y=\Psi_y\hbox{ , }\Phi_2^y=\pm\pi -\Psi_y\hbox{ , }
\Phi_3^y=-\Psi_y\hbox{ , }\Phi_4^y=\mp\pi +\Psi_y .
\end{equation}
Recalling our topological constraint, \Eqnref{topologicalconstraint}, we find 
that it is satisfied as
\begin{equation}
\Phi_1^y+\Phi_2^y-\Phi_3^y-\Phi_4^y=2\pi n,
\end{equation}
with $n=\pm 1$, which is what we expect for an antiferromagnet. Which sign
that applies to a certain plaquette depends on the sublattice associated with
the plaquette as well as the spin-configuration. \Figref{nnngauge} defines
a gauge by introducing the phases $\alpha_y$ and $\beta_y$. Using 
\Eqnref{subfluxes} it is easy to read off the parameters $\alpha_y$ and 
$\beta_y$ from \Figref{nnngauge}. Doing this we find $\alpha_y=-\beta_y=\Psi_y$
for odd values of $y$, and $\alpha_y=-\beta_y=\pm\pi -\Psi_y$ for even $y$'s.

As in the nearest-neighbor case, we will work in momentum space in the 
$x$-direction, while keeping the real space description in the $y$-direction. 
We note
from \Figref{nnngauge} that the phases of the links form a staggered structure,
doubling the size of the unit cell, and introducing a scattering between states
of momenta $k$ and $k+\pi$. To deal with this we introduce the same 
two-component wavefunctions as was used in \Secref{uniform}.

Using this model we have investigated how the optimal stripe evolves as the 
second-neighbor hopping amplitude $t'$ is changed from zero. For small values
of $t'$, approximately, 
$-0.3<t'< 0.15$, the structure of the stripe is largely unchanged.
It is still described by $q_x=\pi$ and vanishing $\xi$. The optimal number of 
holes per unit stripe length also remains ($\delta\simeq 0.46$). 
All that happens is basically that
there are small redistributions of the holes within the stripe.

Concerning the tendency to global phase-separation we have considered the
behavior of $E_{\rm dw}(t')/E_{\rm ps}(t')$, i.e. the ratio of the energy per
hole of the domain wall and phase-separated states, respectively. 
We find that this ratio decreases
as we increase $t'$ slightly from zero, and decreases when $t'$ becomes 
negative. This indicates that a negative $t'$ favors the domain wall 
configurations compared to the global phase-separation but, at least for small
$t'$, the domain walls are still unstable against phase-separation.

When $t'$ becomes larger than $0.15$,
the holes diffuse into the antiferromagnet and widens the domain wall.
The optimal number of holes per unit stripe 
length increases. The spin-twist $\xi$ becomes non-zero making
the antiphase boundary wider. Since we are working with a finite width $w$ of
the domain wall, we get problems when the holes start diffusing away from the
center of the domain wall. For our model to be valid, we must require that 
the domain wall is N{\'e}el-ordered and undoped at its vertical 
edges. This is to avoid surface energies when gluing together the
domain wall with a N{\'e}el region.

In the case relevant for the hole-doped cuprate planes, i.e. $t'<0$, we find
that decreasing the value of $t'$ below $-0.3$ keeps the structure 
of the stripe
rather intact in the sense that $\xi$ remains vanishingly small and that the
holes are localized close to the antiphase boundary. The optimal number of 
holes per unit stripe length does however change, it is reduced as $t'$ is 
decreased, e.g. at $t'=-0.5$ we find that the energy per hole is minimized by
$\delta\simeq 0.38$. However, if we further decrease $t'$ the optimal doping
$\delta$ will rise again as we reach $t'\simeq -0.7$. For such large negative
values of $t'$, the holes will spread into the antiferromagnet just as we 
found in the positive $t'$ case. 
Our numerical calculations indicate that
$\xi$ remains small, i.e. the spin-twist still occurs over a very
short distance.



\section{Conclusion}
\label{sec:conclusion}

We have considered an effective version of the two dimensional {\tJ}-model
where the electrons are considered as completely spin-polarized. The effect of 
such a spin texture is to generate a fictitious topological flux through the
lattice. In this paper we have extended the discussion of a previous 
paper~\cite{Ostlund01} concerning the properties of these fluxes. 

Keeping in mind the result of 
Hasegawa {\em et al.}~\cite{Hasegawa89}, where it is shown that the energy of
a free electron gas on a square lattice is minimized when there is one
flux quanta per particle, the possibility of the system prefering a
flux-generating spin-configuration does not seem remote.

To check the above theory we have performed Hartree-Fock mean-field theory
calculations for the system. In the Hartree approximation it seems like the 
system prefers to form a flux-phase for certain choices of doping and 
coupling constants. However when the exchange-terms are included this effect
seems to vanish and the coplanar spiral phase is energetically more favorable
than the flux phase. Introducing a nearest-neighbor Coulomb repulsion it is 
possible to make the flux-phase energetically most favorable also with the
exchange terms present. However, the calculations indicate that for a wide
range of dopings, these uniform phases are unstable against phase-separation
into an undoped antiferromagnet and a highly doped coplanar spiral phase.
Thus we have to conclude that from a thermodynamic point of view,
we do not expect to find a flux-phase in the phase-diagram of the model we
have considered.

The main part of the paper has been concerned with the generalization of this
construction to describe stripes, directed along either
the (10)- or (11)-direction of the lattice. These stripes are 
appealing as they provide a smooth realization of an antiphase
domain wall, continuously merging two N{\'e}el-ordered regions with opposite 
signs of the order-parameter. The holes naturally
reside within this domain-wall which generates a fictitious flux which 
further can reduce the energy of the holes in domain wall.

Using $J=0.40$, we find that the optimal antiphase domain wall
is site-centered and runs along the $(10)$-direction of the lattice. The
structure of this domain wall is such that there is a sharp spin-twist,
basically only affecting a single row of spins, which aligns the spins 
ferromagnetically in a one-dimensional channel. The doping of the domain wall
is approximately $\delta =1/2$ holes per unit length of the stripe. 
All holes are tightly bound to the domain wall, spreading over three rows of
lattice sites. Due to the narrow spin-structure of this stripe, we found that
the system does not exploit the $\pi$-flux generated through each plaquette by
this domain wall configuration.

We also find that stripes directed along the $(11)$-direction are energetically
very close the optimal $(10)$-stripe, and if we decrease $J$ they will become
the most advantageous domain walls at $J\simeq 0.36$, while increasing $J$
favors the $(10)$-stripe. Looking at the structure of the optimal 
$(11)$-stripes we found that they do not generate any fictitious flux.

Comparing the energetics of the domain walls with global 
phase-separation, we find that within our approximation the global 
phase-separation is favorable. Furthermore, we have incorporated 
second-neighbor hopping in the case of $(10)$-stripes. For small values of 
this 
hopping, the structure of the optimal domain wall remains, while at larger
values the holes starts spreading out, widening the domain wall. 

There are a number of questions that are left unanswered at this point, and
which we believe are interesting to further investigate. Concerning the
next-nearest-neighbor hopping it would be interesting to investigate more
through how it affects the tendency towards global phase-separation, 
and if it can stabilize the stripes. Furthermore, it would be interesting to 
further examine the effect of this hopping on the structure of the stripes,
and also investigate its effects on the $(11)$-stripes.




\end{multicols}
\end{document}